\documentclass[prb,noshowpacs,superscriptaddress,twocolumn]{revtex4-1}

\usepackage{graphicx, subfigure}

\usepackage{amsmath}
\usepackage{amsfonts}
\usepackage{graphicx}

\usepackage{natbib}
\usepackage[final=true]{hyperref}

\usepackage{color}
\usepackage[usenames,dvipsnames]{xcolor}

\usepackage{chngcntr}
\usepackage[shortlabels]{enumitem}

\begin{document}
\title{2D Quantum Turbulence in Polariton Condensates}

\author{S.~V.~Koniakhin}
\email{kon@mail.ioffe.ru}
\affiliation{Institut Pascal, PHOTON-N2, Universit\'e Clermont Auvergne, CNRS, SIGMA Clermont, Institut Pascal, F-63000 Clermont-Ferrand, France}
\affiliation{St. Petersburg Academic University - Nanotechnology Research and Education Centre of the Russian Academy of Sciences, 194021 St. Petersburg, Russia}

\author{O.~Bleu}
\affiliation{Institut Pascal, PHOTON-N2, Universit\'e Clermont Auvergne, CNRS, SIGMA Clermont, Institut Pascal, F-63000 Clermont-Ferrand, France}

\author{G.~Malpuech}
\affiliation{Institut Pascal, PHOTON-N2, Universit\'e Clermont Auvergne, CNRS, SIGMA Clermont, Institut Pascal, F-63000 Clermont-Ferrand, France}

\author{D.~D.~Solnyshkov}
\affiliation{Institut Pascal, PHOTON-N2, Universit\'e Clermont Auvergne, CNRS, SIGMA Clermont, Institut Pascal, F-63000 Clermont-Ferrand, France}

\begin{abstract}
The coexistence of the energy and enstrophy cascades in 2D quantum turbulence is one of the important open questions in the studies of quantum fluids. Here, we show that polariton condensates are particularly suitable for the possible observation of scaling on sufficiently large scales. The shape of raw energy spectra depends on the procedure of condensate excitation (stirring), but the energy spectra of clustered vortices always exhibit the -5/3 power law. In the optimal case, the cascade can be observed over almost 2 decades.
\end{abstract}

\maketitle

\section{Introduction}

Turbulence is a peculiar kind of stochastic behavior with an emergent order, characterized by the formation of self-similar fractal structures at different scales:
\begin{verse}
Big whirls have little whirls,\\
That feed on their velocity;\\
And little whirls have lesser whirls,\\
And so on to viscosity\cite{richardson2007weather}.
\end{verse}
Soon after the famous work of Kolmogorov \cite{kolmogorov1941local}, which determined the scaling $-5/3$ of a direct energy cascade with energy flowing from the injection scale towards smaller scales where it dissipates\cite{obukhov1941spectral,barenblatt1999kolmogorov}, another seminal work by Kraichnan \cite{Kraichnan1967} predicted the possibility of the formation of an inverse cascade in 2D fluids, where the energy is transferred towards larger scales with the associated self-organization of spatial patterns, similar to the formation of the Benard cells\cite{Benard1900}. This result, confirmed numerically\cite{Lilly1972} and experimentally\cite{Sommeria1986} in classical fluids, is based on the mutual incompatibility of the scalings of the cascades associated with 2 conserved quantities: the energy and the squared vorticity (enstrophy), which therefore have to be transferred in opposite directions.

The existence of a similar inverse cascade, suggested for quantum 2D turbulence \cite{Horng2009}, is actually still a matter of a strong debate\cite{tsubota2017numerical}. Contrary to the 3D quantum turbulence, observed in liquid helium \cite{vinen1957mutual,skrbek2011quantum,barenghi2014introduction} and atomic  condensates\cite{tsubota2013quantum,allen2014quantum,stagg2015generation,white2014vortices,Navon2016}, the inverse cascade of 2D quantum turbulence remains elusive even in numerical simulations, let alone real experiments. Indeed, while several works\cite{Horng2009,reeves2013inverse2} report the numerical observation of an inverse cascade with a scaling of $-5/3$, others argue against it \cite{Numasato2010}. The enstrophy in quantum fluids is proportional to the total number of quantum vortices, which can appear and (most importantly) disappear only in pairs.  It is argued \cite{Numasato2010} that the dissipation of enstrophy in quantum fluids could be expected to occur differently from the classical ones: instead of requiring a transfer to smallest scales, it could on the contrary be dissipated at any scale above the vortex size (healing length). For example, two very large clusters rotating in opposite directions and forming a dipole could dissipate vorticity along their mutual boundary, without requiring any transfer to smaller scales associated with the redistribution of vortices and formation of smaller clusters and isolated vortex pairs. So, the enstrophy cannot be \emph{a priori} considered as a conserved quantity which is transferred over scales in order to be dissipated at the smallest ones, and thus the incompatibility of the scaling of  cascades cannot be used to prove the existence of the inverse energy cascade.

Not only the conclusions of the scaling arguments are controversial, but the mathematical limits, imposed on numerical simulations by the properties of the real systems are so stringent, that they prevent one from drawing definite conclusions from the numerically observed energy cascades published in the most recent works. Indeed, one never observes a cascade over more than 1 decade of wave vectors in such simulations (and even in recent experiments with 3D condensates\cite{Navon2016}), and the suggested scaling is usually not a \emph{fit} of the spectral density, but only a guide for the eyes. Actually, since the spectral energy density often presents a transition between large and small scales (either at the injection scale or at the vortex size), any scaling exponent can be suggested as a tangent to such bell-like curve, and the interpretation is  therefore highly arbitrary.

The recent progresses in semiconductor heterostructure manufacturing and spectroscopy techniques \cite{kavokin2003cavity,richard2005experimental,wertz2010spontaneous} make the polariton condensates, formed in microcavities in the regime of strong light-matter coupling, a perfect playground for the experimental studies of Bose-Einstein condensates and associated phenomena like quantum turbulence \cite{Berloff2010}. The coherent propagation of a polariton quantum fluid has been observed at the scale $L$ of hundreds of micrometers \cite{wertz2010spontaneous,Hivet2012}, and the expected coherence decay due to quantum and thermal fluctuations is of the order of several millimeters \cite{Solnyshkov2014} defining the lower cascade bound $1/L$. The upper cascade bound $1/\xi$ is defined by another important condensate parameter, namely the characteristic healing length $\xi$ -- the size of a quantum vortex.
In the numerous experimental observations of quantum vortices  in polariton condensates \cite{Lagoudakis2008,Lagoudakis2011,dominici2018interactions,caputo2018topological} this parameter was estimated to be of the order of a micron. This allows in principle to hope for the observation of an energy cascade in a well-developed turbulence over at least 2 decades of wave vector magnitude. Optical techniques allow studying the cascade in polariton condensates either using interferential analysis \cite{sanvitto2011all,amo2011polariton} of spatial images generated in single-shot experiments\cite{Baumberg2008,Bobrovska2018,Estrecho2018}, or by direct energy spectrum analysis based on simpler to implement angle-resolved photoluminescence (PL) spectroscopy experiments.

The procedure of vortex creation is very important for the  observation of the cascade both theoretically and experimentally. Previously, the schemes of "spoon"-stirring \cite{reeves2012classical,reeves2013inversePhD,skaugen2016vortex} or flowing of the condensate around a set of stationary defects \cite{bradley2012energy,reeves2013inverse2} were used. The artificial vortex imprinting \cite{simula2014emergence,salman2016long} to the condensate was also used.

In this work, we perform an extensive study of quantum turbulence in scalar Bose Einstein condensates, based on the Gross-Pitaevskii equation. We are using physical parameters of an idealized polaritonic condensate, but our results are certainly relevant to other types of quantum fluids. We use the development of modern numerical techniques to maximize the accessible scales. We find that independently of stirring method, the system always contains a mixture of a gas of individual uncorrelated vortices and a fractal structure of vortex clusters. We demonstrate that an $-5/3$ energy cascade can be observed for the clusters, whereas the gas of individual vortices generates a strong signal with a $-1$ power law.  Time-dependent studies of the energy transfer suggest the inverse nature of the $-5/3$ cascade. The corresponding experimental measurements should be possible thanks to single-shot interferometry.

The paper is outlined as follows. In Section II, we present the Methods used in numerical simulations and analysis. In Section III, we present and discuss the results of the simulations. Conclusions are drawn in Section IV. Additional results concerning verification of the numerical procedures are given in the Appendix.

\section{Methods}

The strong coupling of quantized excitonic and photonic modes in a planar microcavity \cite{Microcavities} with one or several quantum wells can be described in the coupled oscillator model with the strong coupling Hamiltonian \cite{Hopfield58}:
\[H = \left( {\begin{array}{*{20}{c}}
{{E_x}}&V\\
V&{{E_c}}
\end{array}} \right)\]
where $E_x=E_{x,0}+\hbar^2 k^2/2m_X$ is the energy of the quantum well exciton, $E_c=E_{c,0}+\hbar^2 k^2/2m_{ph}$ is the energy of the photon in the cavity mode , $k$ is the in-plane wave vector, $m_X\sim m_0$ is the exciton mass, $m_{ph}\sim 3\times 10^{-5}m_0$ is the cavity photon mass ($m_0$ is the free electron mass), and $V$ is the light-matter coupling constant (half of the Rabi splitting).

This coupling gives rise to the anticrossing of the excitonic and photonic modes, and the formation of polariton branches. In the following, we consider only the lower polariton branch in the parabolic approximation. The consequences of such approximation for the simulation of vortices shall be discussed below, together with the other approximations.

\subsection{Simulation}

The basic tool for numerical simulations of an interacting bosonic quantum fluid is
the Gross-Pitaevski equation \cite{bogoliubov1947theory,gross1961structure}. It has been used for the simulation of quantum turbulence in numerous papers \cite{tsubota2017numerical,simula2014emergence,reeves2012classical} including ones devoted to studying the energy cascade\cite{skaugen2016vortex,bradley2012energy,reeves2013inverse2} and behavior of single vortices\cite{flayac2012electric,pigeon2017sustained}. This equation can also be extended, to account for the thermal (uncondensed) part of the fluid \cite{Zaremba1998,Kobayashi2007}, and for other effects, such as the energy relaxation \cite{Pitaevskii58}. However, the description of large-scale systems is difficult to be carried out at the level of full GPE numerical simulation of the quantum fluid, and in this case other models are used, such as the point particle gas approximation with the specific vortex-vortex potentials \cite{valani2018einstein,reeves2017enstrophy}.

The Gross-Pitaevskii equation in dimensionless units reads:
\begin{equation}
    i\frac{\partial \psi}{\partial t}=-\Delta\psi+V\psi+\left(\left|\psi\right|^2-1\right)\psi,
\label{eqn:gp0}
\end{equation}
where $(x,y)=(x_0,y_0)/\xi$ (with healing length $\xi=\hbar/\sqrt{2 g n m}$), $t=t_0gn/\hbar$, $V=V_0/gn$, $\psi=\psi_0/\sqrt{n}$ (the index $0$ marks dimensional variables, $n=|\psi_0|^2$ is the density of the fluid). Having in mind a particular implementation of a quantum fluid based on the exciton-polariton system, we use $m=5\times 10^{-5}m_0$ for the polariton mass (twice the cavity photon mass at zero detuning). $g$ is the strength of the polariton-polariton interaction governed by the exciton-exciton interaction. It can be written as
\begin{equation}
g=6 E_b X_c^2 a_B^2,
\end{equation}
where $E_b$ is an exciton binding energy, $a_B$ is the exciton Bohr radius and $X_c$ is the excitonic fraction. We take the $g$ parameter equal to 5 $\mu$eV$\mu$m$^2$, which coincides with the values given in Ref. \cite{ferrier2011interactions} for GaAs 2D microcavities \cite{PhysRevLett.100.047401}. Operating with densities $n \approx 200~\mu$m$^{-2}$ yields healing length $\xi$ close to $1 \mu$m. A typical time scale for polaritons $t_0=1$~ps corresponds to dimensionless $t=0.9$. Thus, one concludes that micrometers ans picoseconds are quite natural units for consideration of the problem of turbulence in polariton condensates.

In general, we take all parameters corresponding to the state-of-the-art GaAs microcavities, which offer the best performance for the possible observation of the studied effects. However, in order to obtain the clearest possible results on the existence of energy cascades (at least theoretically), we neglect the finite lifetime of polaritons. This is a valid approximation for a real system either if the polariton lifetime is long compared to other time scales, or if the condensate is quasi-adiabatically non-resonantly pumped. The possible side effects of this non-resonant pumping are neglected, as well as the polarization effects, and the non-parabolicity of the polariton dispersion (which could change the $k^{-3}$ spectrum of the vortex core). We also entirely neglect structural disorder effects which in real systems might play an important role in vortex dynamics. All these effects are left for future studies. The choice of the polariton system is important because of the possibility of performing single-shot interference measurements, allowing the detection of the spatial position of vortices, as we discuss below.

In our numerical simulations, the time step was 0.01 ps and an $N \times N = 1024 \times 1024$ mesh was used. The Laplace operator was calculated using the Fourier transform and in time the third order Adams–Bashforth scheme was used. We used \textsc{Matlab} and \textsc{Mathematica} packages for numerical solution of DDGPE and further analysis. The size of the square-shaped space region where the simulation was performed was $L=$1024 $\mu$m, which corresponds to the maximal wave vector $k_{max}=\sqrt{2}\pi N/L \approx 4 \mu$m$^{-1}$. Higher wave vectors are required for a better description of the vortex core.

\subsection{Stirring the condensate}

\begin{figure}[tbp]
  \centering
  \includegraphics[width=0.49\textwidth]{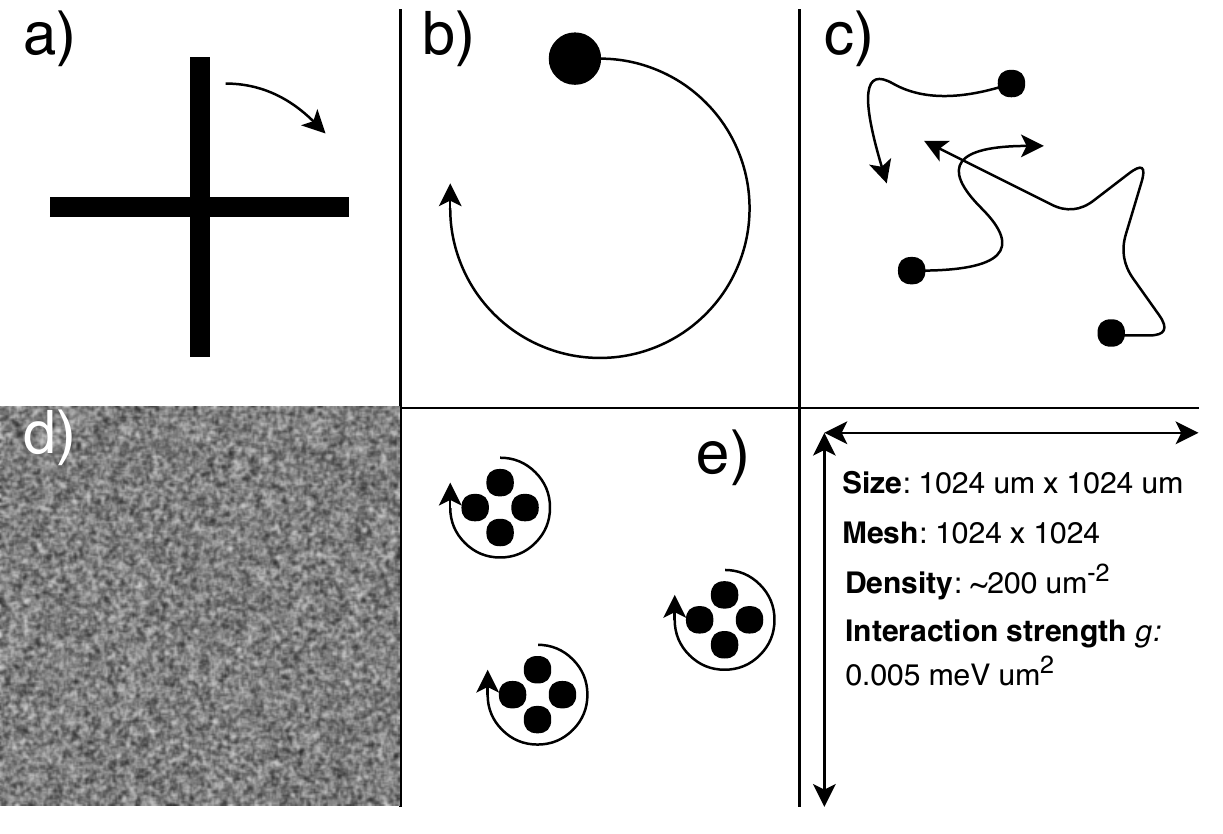}\\
  \caption{Schematic representation of the employed stirring strategies. a) Large cross, b) Classical spoon, c) Brownian spots d) White noise e) Gauss-Laguerre potential}\label{fig1_stirring}
\end{figure}

The main feature of the turbulence is the energy flow from the injection scale towards other scales. It is this flow that leads to the formation of self-similar spatial structures. A cascade should manifest itself in the so-called incompressible energy part, associated with rotation (see below). Thus, the observation of cascades, either direct or inverse, absolutely requires the formation of quantum vortices, and not just of density waves. In classical 2D turbulence, a simplest random-potential scheme has been shown to be sufficiently efficient for the observation of a large-scale inverse energy cascade \cite{Boffetta2000}. In quantum 2D turbulence, such method does not allow to create vortices efficiently, because, contrary to the classical case, creating a pair of well-defined vortices with a vanishing order parameter in their centres requires a finite amount of energy\cite{Jones1982} $E_{pair}\approx 6\hbar^2 n/m$ precisely because these vortices are quantum.


As mentioned above, the first strategy used was the condensate stirring by a propagating potential defect \cite{reeves2012classical} or flow around stationary defects \cite{bradley2012energy,reeves2013inverse2}. Random imprinting of vortices followed by healing by simulation with imaginary time has also been used\cite{salman2016long,simula2014emergence}. In polariton condensates, persistent vortices have already been shown to appear because of the flow of the condensate against a random potential \cite{Sanvitto2010}.

In the present manuscript, we have compared several different stirring strategies  (see Fig. \ref{fig1_stirring}):
\begin{enumerate}[a)]
    \item Large cross-like potential
    \item Classical rotating spoon
    \item Several spots in brownian motion
    \item White noise with spatial correlations (for comparison with a classical fluid \cite{Boffetta2000})
    \item Several small potential wells defined by the intensity of the interference of 2 Gauss-Laguerre (GL) beams
\end{enumerate}

As we shall discuss below, these procedures inject energy at different scales. To obtain a quasistationary configuration, we have used very long times for the stirring of the condensate (5 ns), for its relaxation (20 ns), and for the averaging during the extraction of the cascade (5 ns). However, the analysis of the dynamics presented in the final part of the work demonstrates that the characteristic formation time of the cascade is of the order of 200 ps, which is much closer to the lifetimes of the state-of-the-art cavities.

\subsection{Extraction of the energy spectra}

The Kolmogorov energy cascade is expected to form in the Incompressible Kinetic Energy  (IKE) spectral density, and its observation requires the separation of the density-weighted velocity field into the compressible and incompressible part \cite{bradley2012energy, kowal2010velocity}, with the selection of the latter. Importantly, the spectral energy density of the incompressible part of the velocity field can also be calculated analytically from the positions and the signs of the quantum vortices \cite{bradley2012energy,skaugen2016vortex}. This is possible because the quantum fluid is irrotational and all the vorticity is concentrated only in vortices.  One also needs such parameters of condensate as its density $n$, interaction constant $g$ and polariton mass $m$. One writes the IKE spectral density (IKE spectrum) as

\begin{equation}
    E^{(i)}(k) = N_{\mathrm{vort}} \Omega \xi^3 F(k\xi) G(k),
\end{equation}
where $F=\Lambda^{-1}f(k\xi\Lambda^{-1})$ is the single vortex spectrum,  $N_{\mathrm{vort}}$ is the total amount of vortices, $\Omega = 2\pi \hbar^2 n / (m\xi^2)$ is the ensthropy quantum, the parameter $\Lambda=0.8249...$ and the function $f(z)$ writes 

\begin{equation}
    f(z)=\frac{z}{4}\left( I_1\left(\frac{z}{2}\right) K_0\left(\frac{z}{2}\right) - I_0\left(\frac{z}{2}\right) K_1\left(\frac{z}{2}\right)   \right).
\end{equation}

The function $G(k)$ is shaped by the coordinates $\mathbf{r}_{i,j}$ and the signs $\kappa_{i,j}$ of the vortices:
\begin{equation}
    G(k)=1+\frac{2}{N_{\mathrm{vort}}}\sum_{i=1}^{N_{\mathrm{vort}}-1} \sum_{j=i+1}^{N_{\mathrm{vort}}} \kappa_i \kappa_j J_0 \left( k|\mathbf{r}_i-\mathbf{r}_j| \right),
\end{equation}
where the indices $i$ and $j$ enumerate all vortices.

This approach allows not only to find the total incompressible energy spectrum, but also to consider the contributions of single vortices and clusters separately \cite{valani2018einstein,skaugen2016vortex}, which turns out to be important in order to observe the Kolmogorov cascade at a large scale. The cluster selection algorithm was adopted from Refs. \cite{reeves2012classical,valani2018einstein} with an additional optimization available in \textsc{Mathematica}. The positions of vortices were determined from the phase of the wave function.

More details on the numerical methods can be found in the Appendix.


\section{Results}

In this section, we first study the total incompressible energy spectra for different stirring strategies. We show that, independently of the stirring strategy, such spectra are strongly dominated by the contribution of individual vortices, which prevents the observation of the cascade over large energy scales. In the second subsection, we remove the contribution of individual vortices, keeping only the one belonging to clusters.  This procedure reveals the cascade over a much large range which confirms the arrangement of clustered vortices in large scale self-similar structures. In the last subsection, we demonstrate the inverse nature of the cascade by analyzing the time evolution of the energy distribution during the stirring procedure.

\subsection{Total IKE spectra}

\begin{figure}[tbp]
  \centering
  \includegraphics[width=0.49\textwidth]{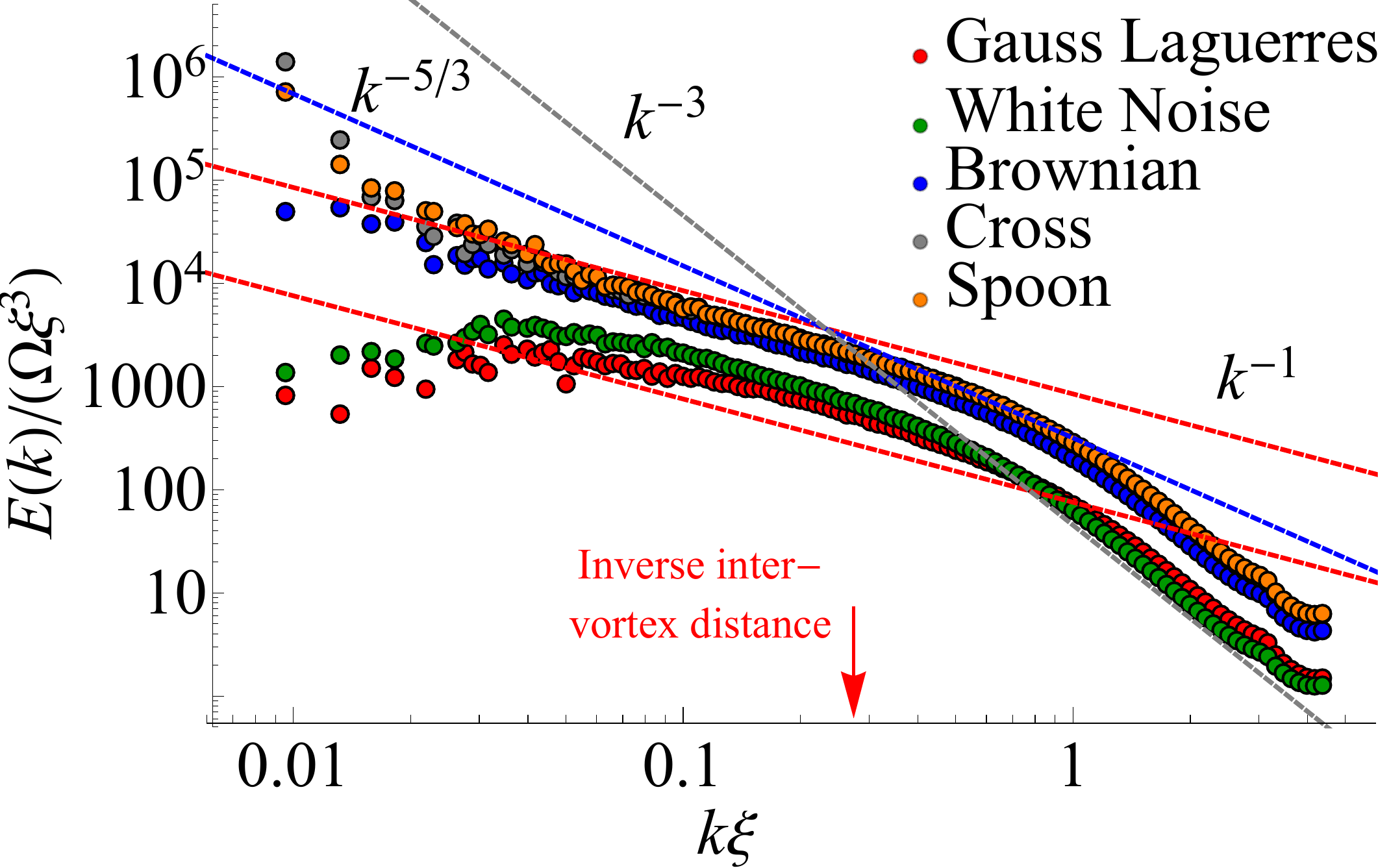}\\
  \caption{The IKE spectra obtained numerically by decomposition in the reciprocal space for different stirring strategies. The red arrow is for the inverse intervortex distance $k_l$.}\label{fig1_numerical}
\end{figure}

Figure~\ref{fig1_numerical} shows the IKE spectra obtained for various stirring strategies. The incompressible energy part was separated by decomposition in the reciprocal space (see Methods).
One can expect to observe the $-5/3$ power law cascade in the IKE spectrum only between the wave vectors $k_L=2\pi/L$ ($L$ is the system size) and $k_l = 2\pi/l$, where $l$ is the mean inter-vortex distance ($l$ was approximately 20~$\mu$m in most number of our simulations and thus $k_l\xi\approx 0.3$). In Fig.~\ref{fig1_numerical}, such power law is visible only for the cross and spoon stirring, and only in a narrow wave vector range (in the vicinity of $k\xi=0.02$). The difficulty to observe the $-5/3$ power law characteristic for the formation of multiscale structures is explained by the large contribution of single vortices, as we show in the next section. 

It is natural\cite{bradley2012energy} to measure the IKE spectral density in the units of ensthropy $\Omega \xi^3$. According to the definition, the function $F(k\xi)$ is of the order of 1 at $k\xi=1$ and the function $G(k) \approx 1$ at $k \gg k_l$. Thus, in the vicinity of the point $k\xi = 1$ the magnitude of IKE spectrum estimates the total amount of vortices in the system $N_{\mathrm{vort}}$.


For $k\xi$ between $k_l\xi$ and 1 one should observe the energy spectrum of a single vortex: $k^{-1}$, because at this scale a vortex does not have any neighbors to form any structures. For wave vectors larger than $1/\xi$ (short length scale), one obligatory observes a $k^{-3}$ law which is a fingerprint of the vortex core wave function (this might be different for exciton-polaritons in some regimes because of their non-parabolic dispersion that we neglect here). In Fig.~\ref{fig1_numerical}, all stirring strategies exhibit similar behavior at wave vectors higher than $k_l$ : there is a $k^{-1}$ power law below $1/\xi$ and $k^{-3}$ above $1/\xi$. 

It is interesting to note that in Ref. \cite{skaugen2016vortex} the $-5/3$ cascade signatures have been observed between $k_l$ and $1/\xi$. In Refs. \cite{bradley2012energy,reeves2012classical} there was also no transitional $k^{-1}$ regime between $1/\xi$ and $k_l$, and the $-5/3$ cascade started immediately after $1/\xi$. The absence of an intermediate  region with $k^{-1}$ power law in these works might be explained by the short inter-vortex distance which is close to the healing length, or by a large variation in the intervortex distance.

Finally, the differences between stirring strategies can be observed in Fig.~\ref{fig1_numerical} at $k\xi$ smaller than 0.02-0.03. Indeed, the stirring based on the classical spoon and large cross generates large-scale vortex clusters. The energy injection for these strategies is still efficient at the scales of  $k\xi= 0.01$. The three other stirring procedures do not inject energy at large scales and the IKE spectra drop  below $k\xi = 0.03$. However, in all cases most of the energy spectrum is dominated by the signal arising from single vortices, which strongly hinders the observation of the Kolmogorov energy cascade because of the wide spreading of single vortex energy in k-space. So in the next section, we are going to change our treatment procedure using real space selection allowing to eliminate single vortices in order to keep only the part of the IKE located in clusters.

\subsection{IKE spectra of clustered vortices}

\begin{figure}[tbp]
  \centering
  \includegraphics[width=0.49\textwidth]{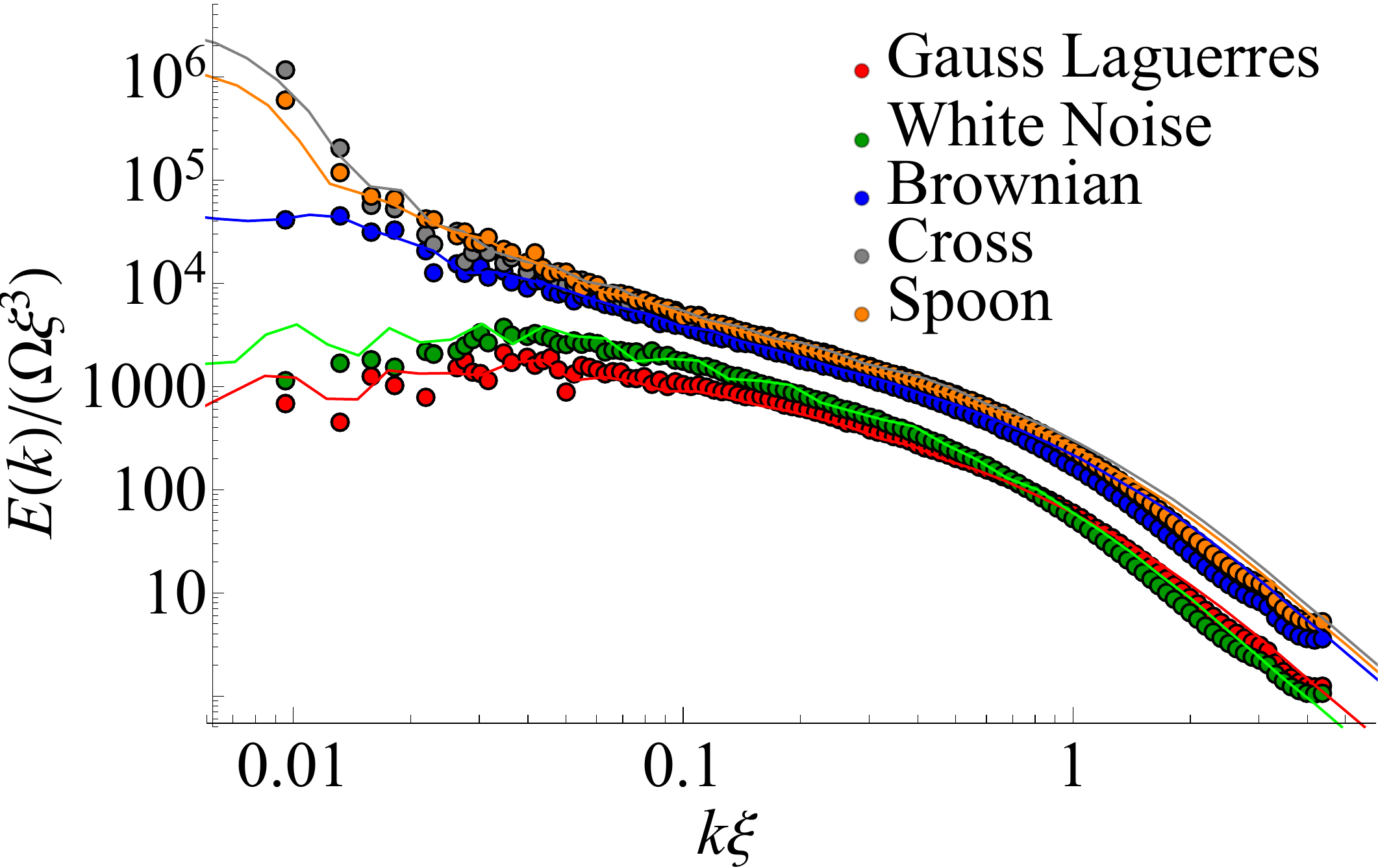}\\
  \caption{IKE spectra obtained numerically (points) and analytically for all vortices (solid lines). In the range from 0.02 to 1 $k\xi$ both methods give the same result.}\label{figA1}
\end{figure}

We are going to compute the IKE using the analytical procedure described in the Methods section and in the Appendix. This procedure is based on the detection of vortices in real space. Indeed, the knowledge of the wave function at any time allows to determine the position of all vortices. The velocity field induced by these vortices is the whole incompressible velocity field. Once the vortex position is known, the incompressible velocity field can be computed analytically (see Methods).
Fig. \ref{figA1} shows the result of this procedure (for all stirring procedures). These results are compared with those obtained in the previous section based on a decomposition in reciprocal space. One can indeed check that both methods coincide. 

It is then possible to make one step further by determining if a given vortex is single, in a dipole, or in a cluster (details on the procedure are given in the Appendix). This is illustrated by Fig.~\ref{fig4_phaseBR} showing a snapshot of the phase of the condensate stirred by Brownian potentials. The winding of each vortex is shown by colour ($+1$ - red, $-1$ - blue). Single and dipole vortices are marked by small circles. Vortices belonging to clusters are marked by large circles. One clearly sees that a large fraction of vortices (about 50\%) belong to clusters. At the same time, it is natural that the signal from the other 50\% that are not in clusters is quite important in the total IKE spectrum.

\begin{figure}[tbp]
  \centering
  \includegraphics[width=0.49\textwidth]{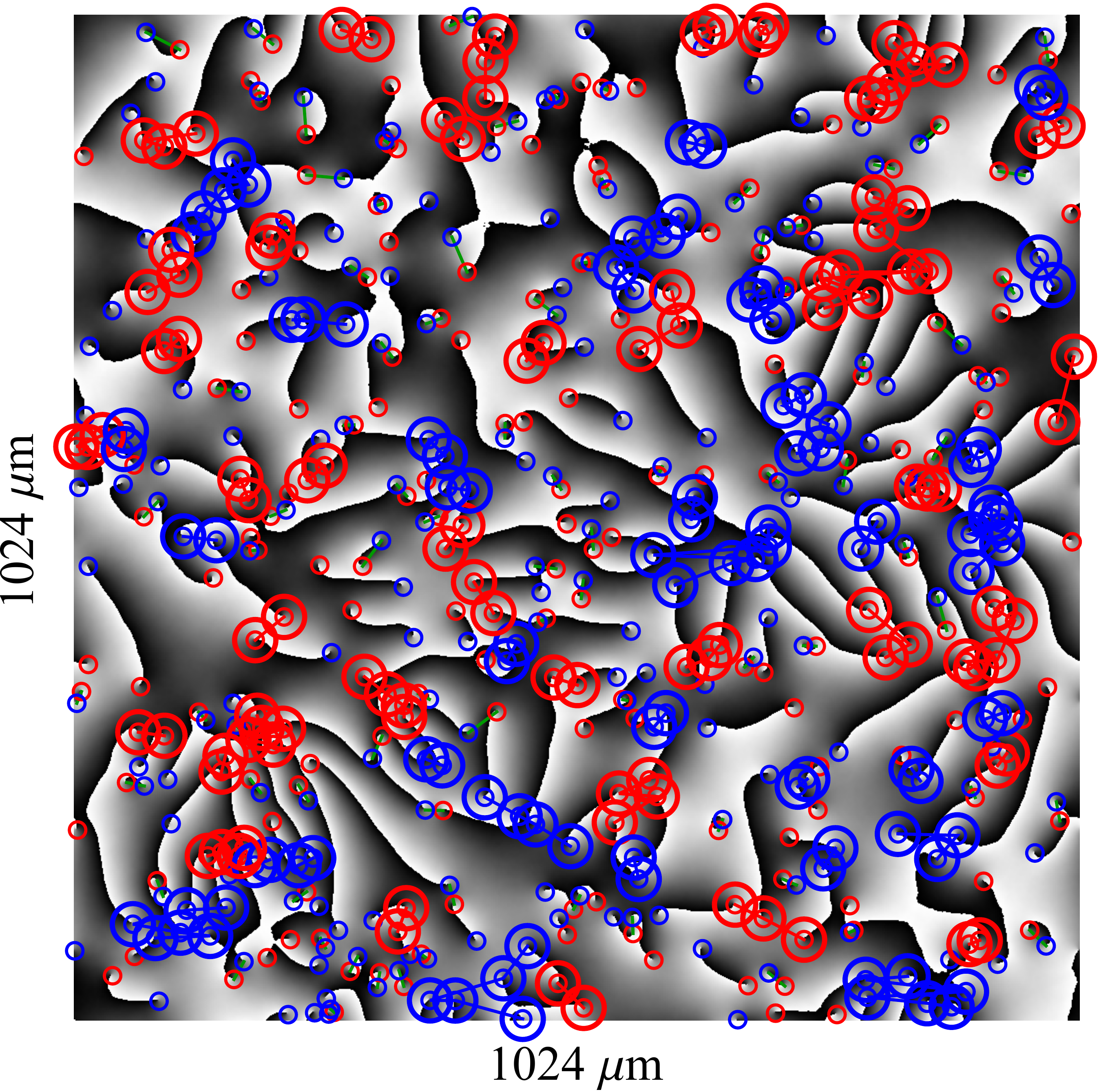}\\
  \caption{Phase of the wave function for Brownian stirrers combined with the results of clustering procedure. The vortices belonging to the clusters are highlighted with the large circles. Relatively rare vortex dipoles are connected by green lines.}\label{fig4_phaseBR}
\end{figure}

\begin{figure}[tbp]
  \centering
  \includegraphics[width=0.49\textwidth]{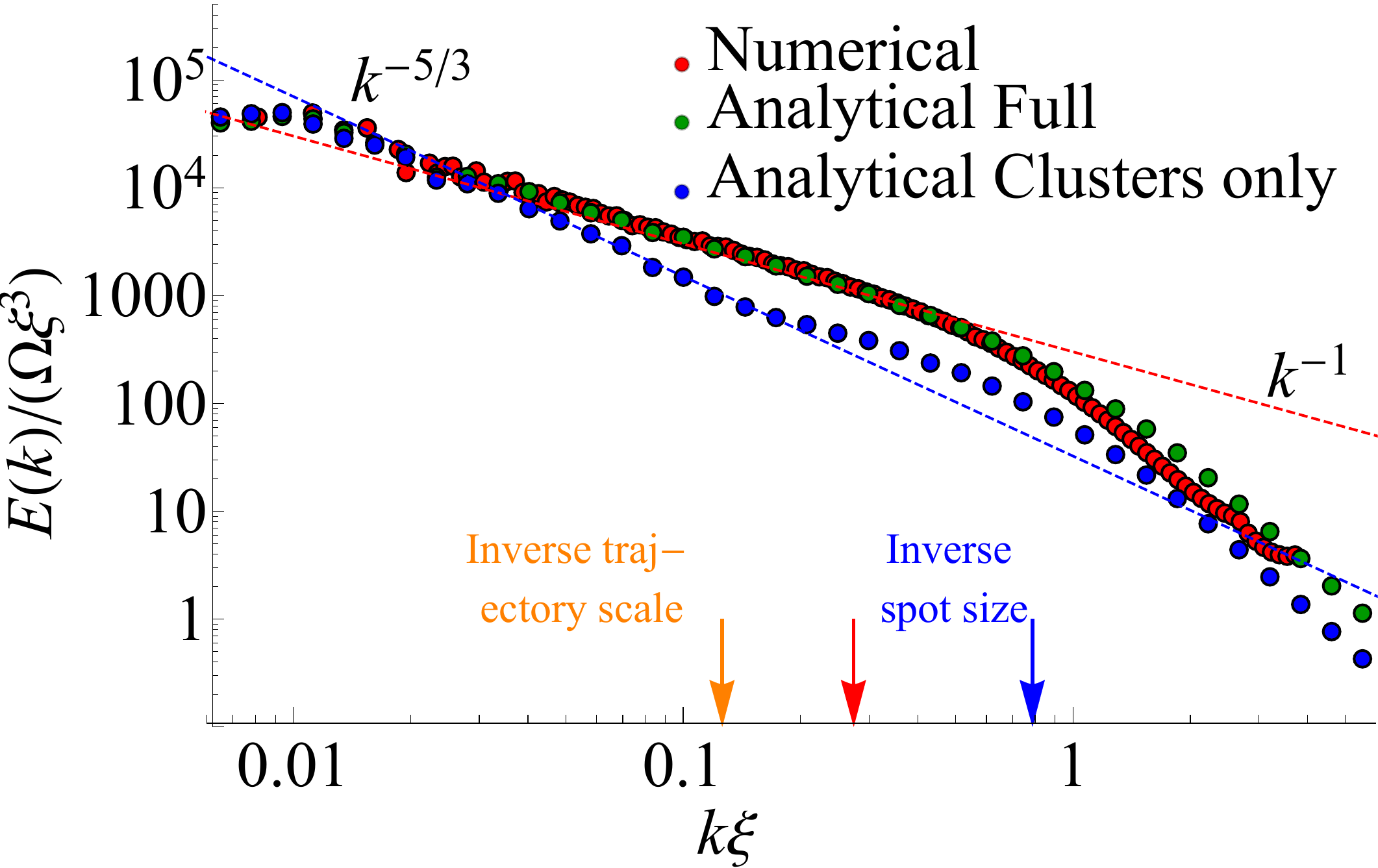}\\
  \caption{IKE spectra for Brownian stirrers. Red dots are for numerical procedure, green dots are for analytical procedure for all vortices, blue dots are for clustered vortices only. The dashed lines are the guides for the eye with $-5/3$ (blue dashed) and $-1$ (red dashed) power functions. The blue arrow shows the inverse potential spot size ($\sim 7\xi$) and the orange one shows the inverse characteristic scale of Brownian motion ($\sim 50 \xi$). The red arrow is for $k_l$.}\label{fig5_BRdetailed}
\end{figure}

Figure~\ref{fig5_BRdetailed} compares the IKE spectra with and without selection of clustered vortices. A $-5/3$ power law is visible in both cases between $k\xi=0.01$ and $k\xi=0.05$. The spectra differ above $k\xi>0.05$. The removal of the single vortex and dipole contributions reveals a very clear $-5/3$ slope over more than one order of magnitude, which was hidden in the total IKE spectra. It becomes clear therefore, that the removal of single vortices is crucial for the analysis of the turbulence phenomena via the incompressible energy spectrum.

\begin{figure}[tbp]
  \centering
  \includegraphics[width=0.49\textwidth]{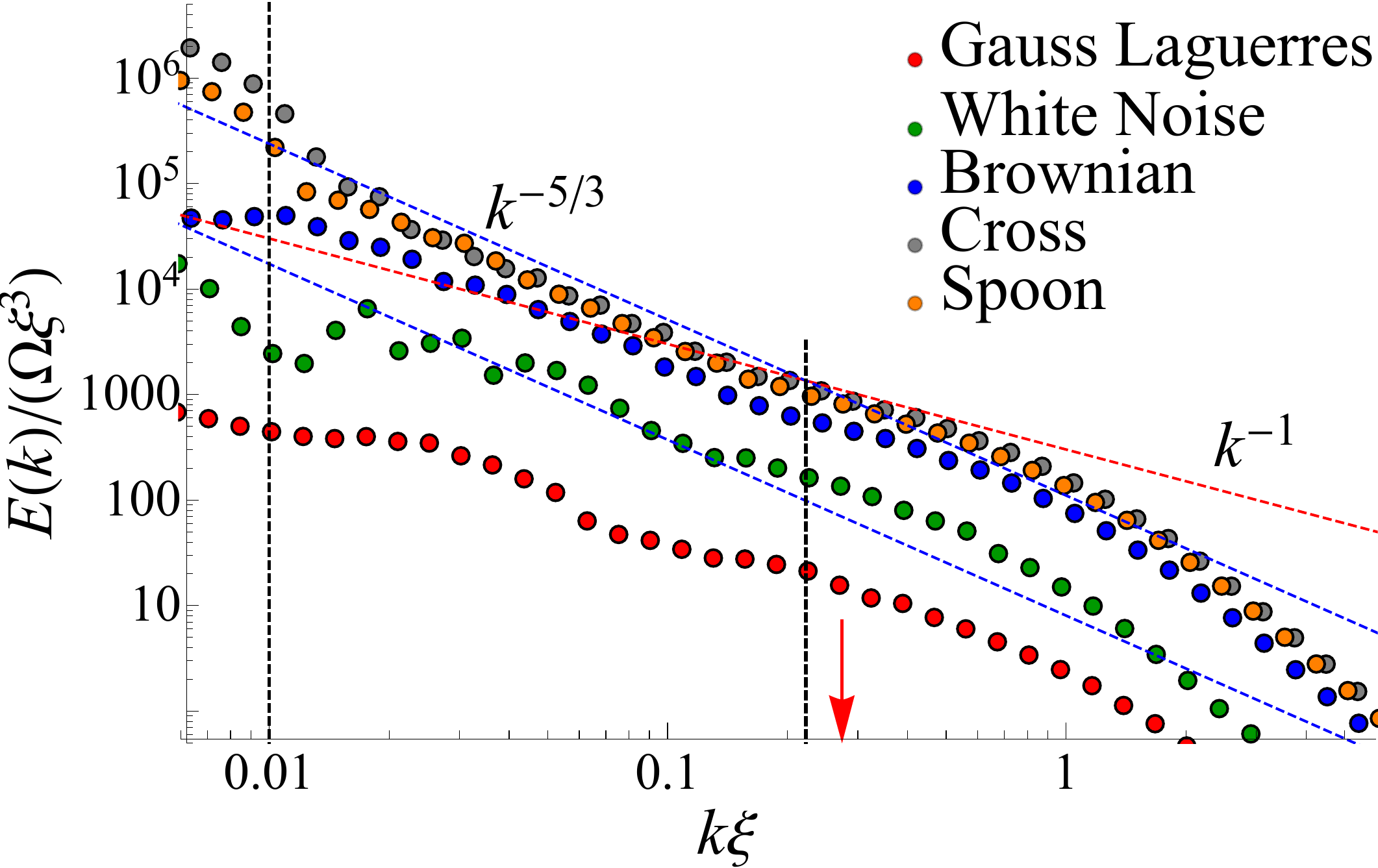}\\
  \caption{The analytically obtained IKE spectra for all 5 stirring strategies for the clustered vortices only. The red arrow is for inverse intervortex distance $k_l$. The vertical lines cut the range where -5/3 power law is observed.}\label{fig3_OnlyClusters}
\end{figure}

The power spectra computed with the same spatial selection procedure for all five stirring procedures are shown on Fig. \ref{fig3_OnlyClusters}. A $-5/3$ power law is now visible in all cases, and also extends over more than one order of magnitude over $k$. This is demonstrating the presence of self similar structures of vortices with their size varying from about 30~$\mu$m to 600~$\mu$m.

\begin{figure}[tbp]
  \centering
  \includegraphics[width=1\linewidth]{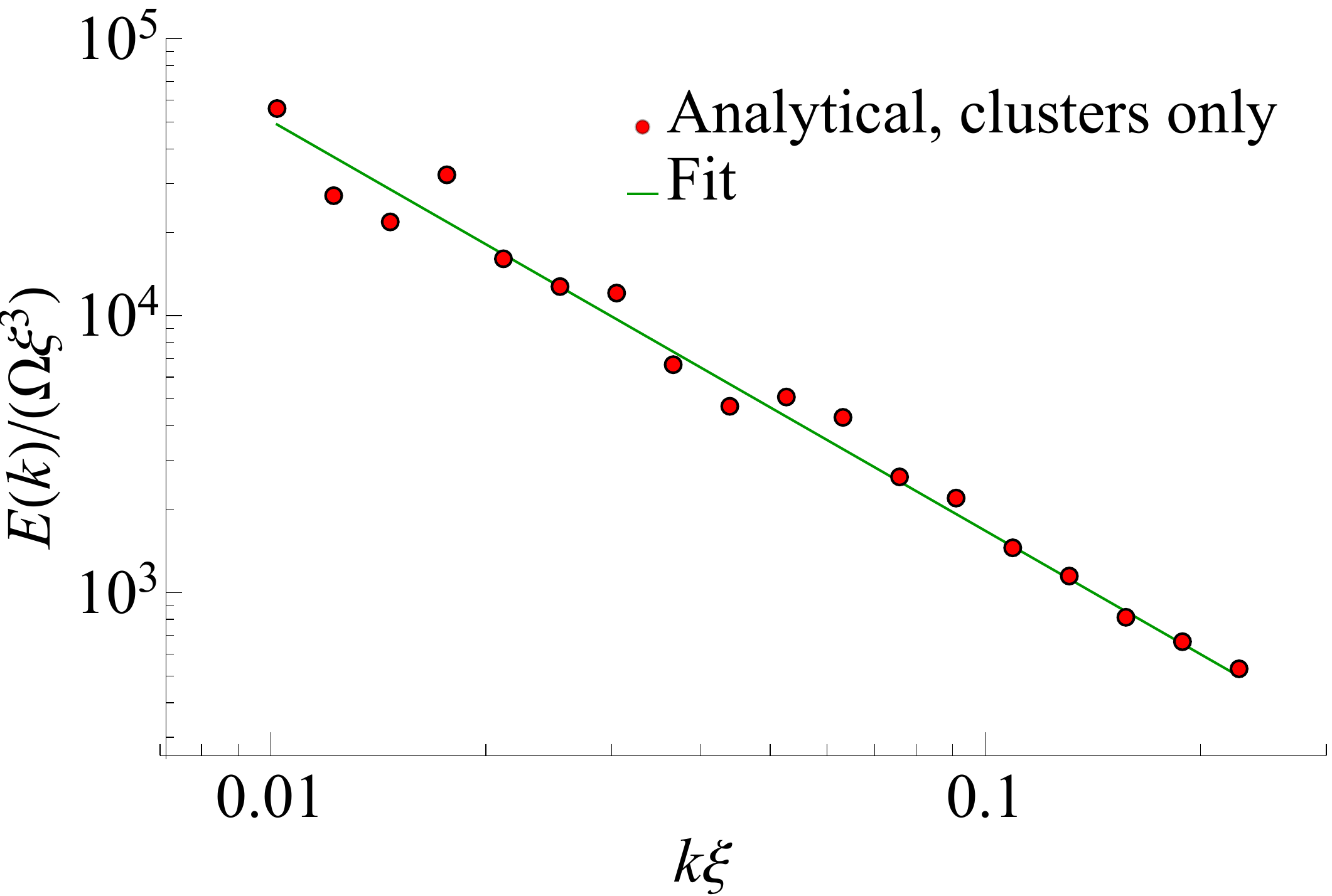}\\
  \caption{IKE spectrum for Brownian stirrers (clusters only) with a power-law fit giving $\gamma=-1.5\pm0.2$ (expected $-1.66$).}\label{fig6b}
\end{figure}

In order to check that the observed power law indeed corresponds to the expected scaling of $-5/3$, we fit the IKE resulting from the analytical procedure with an allometric (power) function $f=ax^\gamma$ with fitting parameters $a$ and $\gamma$ (Fig.~\ref{fig6b}, dots and solid line). We use the non-linear least squares procedure with the Levenberg-Marquardt error minimization algorithm, with the confidence interval for parameter values obtained from the variance-covariance matrix using the asymptotic symmetry method. The fit shows that the expected value $-1.(6)$ is within the bounds of the confidence interval: $\gamma=-1.5\pm 0.2$, confirming the presence of the Kolmogorov scaling over more than one order of magnitude of wave vectors and energies. We stress that although the precision is relatively low, this is a true fit of the numerical experiment, and not just a guide for the eyes. The importance of performing a fit is underlined by the fact that even for a completely random arrangement of vortices (vortex gas) obtained without solving the Gross-Pitaevskii equation and so without any possible self-organization effects linked with quantum turbulence, the IKE spectrum naturally demonstrates a bell-like curve, which can have a tangent slope of $-5/3$ in a certain region. Thus, a thorough analysis confirming the existence of a large scale cascade is really required to draw any conclusions on the quantum turbulence.

Analyzing power law dependencies can be particularly difficult, because one needs to confirm that the observed approximately linear distribution on the loglog plot is best explained by a power law \cite{Clauset2009}, and not by some other distribution function (for example, exponential or log-normal). When there are no other means, one has to check if the distance of the measured distribution from the ideal one is not higher than for simulated power-law distributions. In a particular physical system, however, the power law energy distribution arises from the formation of the self-similar spatial structures, that is, fractal clusters. In the next section, we analyze the spatial distribution of vortices in order to confirm that the observed power-law distribution is not accidental.

\subsection{Fractal dimension of vortex clusters}

The Kolmogorov's arguments for the existence of cascade in classical turbulence are based on the self-similarity of observed spatial patterns at different scales. This self-similarity in mathematics is what characterizes fractal structures. An inherent property of fractals is their non-integer dimensionality: a fractal formed of an infinite number of points on a plane is neither a 2D object like a polygone, nor a 0D object like a point, but something between the two. Thus, checking if the clusters of vortices exhibit a non-integer (fractal) dimension, allows us to prove that their spatial patterns are indeed self-similar, as required for the formation of an energy cascade.

\begin{figure}[tbp]
  \centering
  \includegraphics[width=0.49\textwidth]{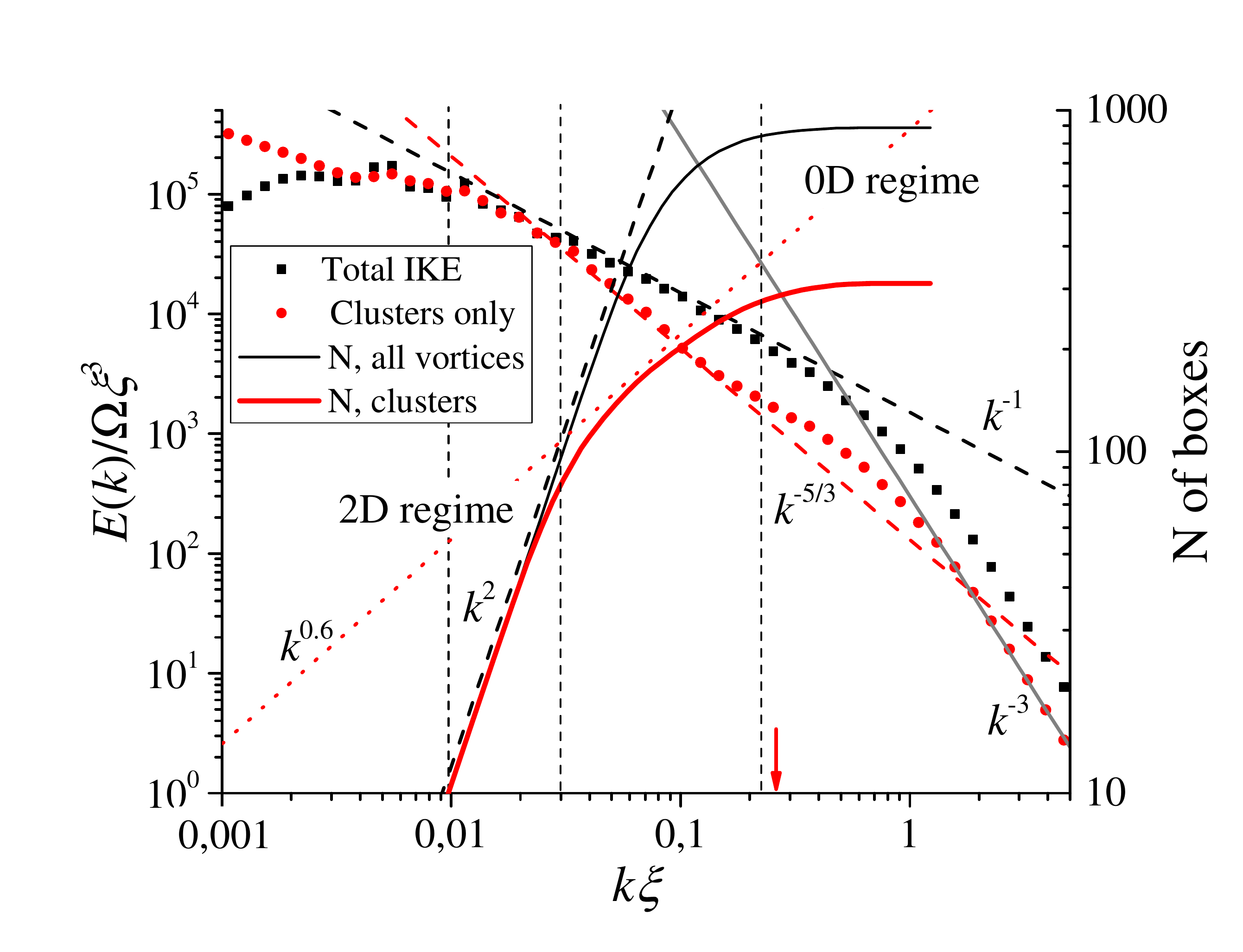}\\
  \caption{Fractal (box-counting) dimension of vortex clusters. Solid curves show the dependence of box counts to cover all vortices on the size of the boxes. Dots are for analytical IKE spectra. Red color is used for clustered vortices and black color is used for all vortices. Dashed lines give the eye guides for some important powers. Red arrow shows the characteristic intervortex distance.}\label{fig_frac}
\end{figure}

One way to obtain the fractal dimension is the box-counting approach, and the corresponding dimension is called box-counting or Minkowski-Bouligand dimension. This approach consists in covering the studied object by a mesh with the cells (boxes) of size $\varepsilon$ and counting the number of boxes required to fully cover the object $N_{box}(\varepsilon)$ for various the mesh sizes $\varepsilon$. For our case, we plot the curve of box count $N_{\rm box}$ to cover all vortices vs. box size $\varepsilon$. The slope of tangent line for this curve finally gives the box-counting dimension of the pattern formed by the clusters of vortices. The results of the analysis are shown in Fig.~\ref{fig_frac}. The box size $\varepsilon$ is given in terms of the corresponding wave vectors $2\pi/\varepsilon$, to have a common horizontal axis with the energy distribution plots. We consider only the best configuration - the case of Brownian stirrer. Vortices were treated as points with coordinates $\bf{r}_i$ obtained from the wave function in the same manner as for analytical calculation of IKE spectra. The system size was 2048~$\mu$m. For this analysis, we were only using vortices of the same sign.

Fig. \ref{fig_frac} shows that box count vs. box size curve for all vortices (without clustering procedure) has a rapid transition from effective zero dimensional structure at large wave vectors (for which a single vortex is a 0D object), to the more or less homogeneous two dimensional structure at large scales (characterized by the power law with exponent 2). Both for all vortices and for the clustered vortices 0D regime starts for wave vectors larger than wave vector of mean intervortex distance $k_l$. The limit of number of boxes in Fig.~\ref{fig_frac} at large wave vectors is straightly the amount of vortices in the system. At low wave vectors, one can estimate $N=(L/\varepsilon)^2$. 

Contrary to the case of all vortices (black), for clustered vortices only (red) one can observe a transitional regime between 2D and 0D limits, with the fractional dimension of the order of 0.6. The size scales at which this regime is present straightly match with the region of $-5/3$ power law in IKE spectra. On the contrary, the rapid transition from 0D to 2D regimes when considering all vortices corresponds to domination of single vortices and vortex pairs in energy storage and thus to the $-1$ power in raw IKE spectra. To check the applicability and robustness of the realization of the used approach, the comparison with a random distribution of points and with an artificially created pattern in the shape of the Sierpinskii triangle have been carried out (see Appendix).

In the next subsection, we study the time evolution of the energy spectra in order to establish the nature (direct or inverse) of the observed energy cascade.

\subsection{Dynamics of the energy redistribution and cascade formation}

In order to understand the formation of a $-5/3$ power law region in the IKE spectra, we analyze the time dynamics during the first stages of stirring by Brownian potentials. In this section, we did not perform the cluster selection procedure in order to keep track of the total IKE spectrum. Similar analysis of the time evolution of the energy distribution has been carried out in previous works \cite{bradley2012energy,reeves2013inverse2}.

\begin{figure}[tbp]
  \centering
  \includegraphics[width=0.49\textwidth]{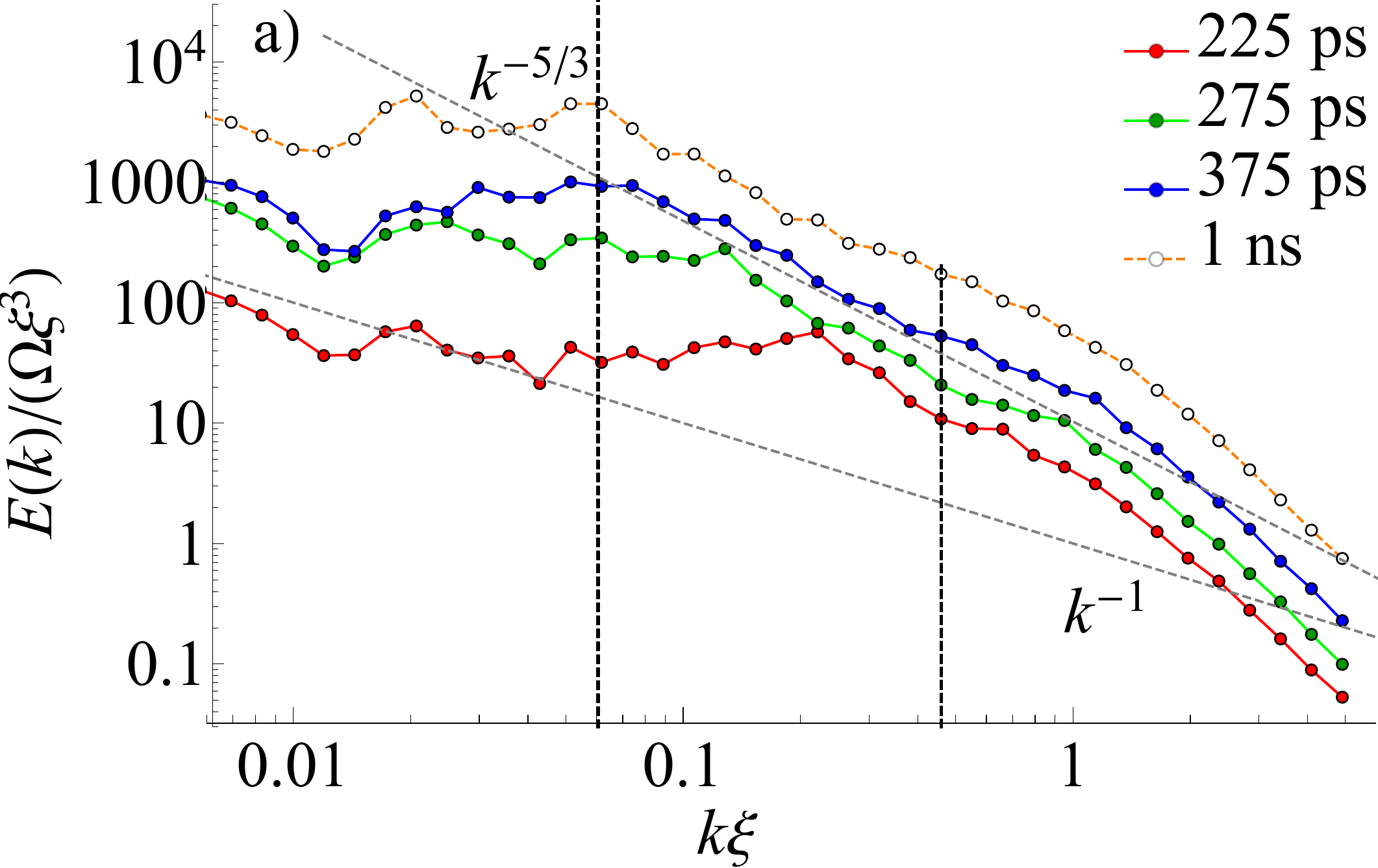}\\
  \includegraphics[width=0.49\textwidth]{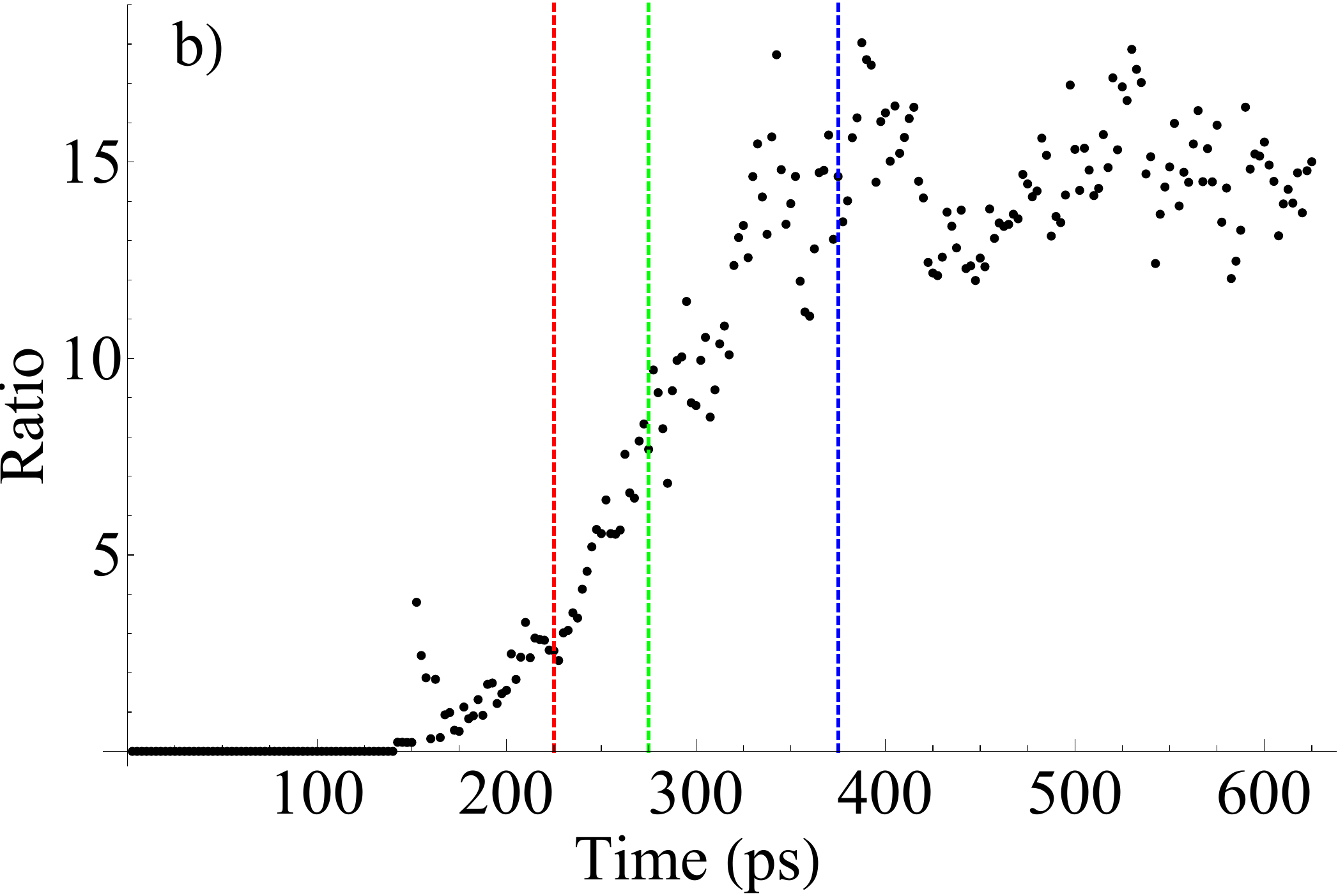}
  \caption{a). Net IKE spectra obtained analytically for Brownian stirrers at several time moments during stirring.  b) Ratio of the spectral energy density at two wave vectors shown by dashed lines in (a), as a function of time. Vertical lines in (b) correspond to the moments of time in (a).}\label{fig5_formation}
\end{figure}

Fig.~\ref{fig5_formation}(a) shows the IKE spectra at four different times after the start of the stirring. The corresponding wave vectors are marked with dashed lines in panel (a).  One sees that at the earlier time (225~ps, red dots) the kinetic energy is mostly concentrated at large wave vectors (small size scales), which corresponds to the injection scale (spot size, peak at approx. $k\xi=0.3$). Then the kinetic energy is transferred from high wave vectors to lower wave vectors versus time.
This is directly visible on the IKE spectra.  It is also quantitatively confirmed in panel (b), showing the ratio of the spectral energy density measured at low and high wave vectors. This ratio grows from 0 at early times, when there is no energy at all at small wave vector, to about 15. One can see that this process takes about 200~ps. This energy redistribution from small scales to large scales due to the intervortex interactions clearly confirms the formation of the inverse Kolmogorov cascade.  The relative rapidity of this process provides an \emph{a posteriori} justification for neglecting the polariton lifetime (which can be of the order of hundreds of ps) in the simulations.
Interestingly, the energy spectrum at 500 ps shows a quite extended $-5/3$ slope without eliminating isolated vortices. This situation corresponds to an optimal moment of time, when the Kolmogorov cascade has built up, while the fraction of individual vortices remains low. The dashed curve with hollow circles shown for comparison in the same figure for the situation at 1~ns demonstrates a growth of the maximum at high wave vectors due  to the single vortices, which leads to the narrowing of the $-5/3$ region. At even later moments of time, when the stirring stops, strong currents break up some of the clusters increasing the relative fraction of individual vortices even more (by up to 20\%). The final conclusion is that at any moment of time removing the contribution of individual vortices allows to increase the scale of the observation of the $-5/3$ cascade.

The difference in the IKE spectra for different stirring strategies stems from the limited efficiency of the energy redistribution at large scales. If the energy is injected at a scale which is too low, then the structures of the largest scales just cannot form, because of the decay of the vorticity at all scales. We note that the observed signature of the inverse cascade does not rule out the presence of a direct one: the energy can be transferred from the injection point in both directions. For us, the most important was to demonstrate the possibility of the inverse cascade, debated for a long time.

\section{Discussion and conclusion}

The existence of the $-5/3$ cascade in 2D quantum turbulence is currently a matter of scientific debate, and its direct observation is quite difficult, even in numerical experiments.
Even with the maximal efficiency of the modern computing hardware (without recurring to supercomputers), we managed to clearly observe and fit the cascade only over 1-2 orders of magnitude. In experiments, obtaining even 1 order of magnitude might be quite challenging. While the scales of the experimental observation in modern microcavities could cover several orders of magnitude in space or wave vector, additional complications arise from the fact that the high-wave vector limit for the cascade is not the healing length $\xi$ (of the order of  1 $\mu$m), but the mean intervortex distance determining $k_l=\frac{2\pi}{l}$. $k_l$ by its nature is greater than $\xi$, and in our study it was typically one order greater: tens of microns.

The numerical IKE spectra are dominated by single vortices with a characteristic  $-1$ slope. Removing these single vortices and vortex dipoles, while keeping the vortex clusters, allows to observe the $-5/3$ IKE spectrum  for all stirring strategies. The usage of Brownian stirrers gives the most extended $-5/3$ region on IKE spectrum after vortex clusters detection procedure. Fitting confirms that the expected scaling falls within the bounds of the confidence interval. We observe a non-integer (fractal) dimension of the vortex clusters at the same scales.

We demonstrate that the observed $-5/3$ is a result of the energy redistribution during the initial moments of stirring. The energy is injected at relatively small scales and transferred to the larger scales (smaller wave vectors). The analysis of the time dependence of the energy stored in large-scale and small-scale structures supports the hypothesis of the inverse energy cascade.

To conclude, the direct observation (using the angle-resolved luminescence detection) of the $-5/3$ cascade in the energy spectrum still remains a challenging task for polaritonic Bose-Einstein condensates. It will require single shot time resolved measurement of the amplitude and phase of the wave function, followed by the clustering procedure. Still, among the different considered stirring procedures the Brownian stirrers are preferable. Time-dependent studies should also allow to observe the energy redistribution during the formation of the cascade.

\begin{acknowledgments}

We acknowledge the support of the ANR "Quantum fluids of Light" project (ANR-16-CE30-0021) and of the ANR program "Investissements d'Avenir" through the IDEX-ISITE initiative 16-IDEX-0001 (CAP 20-25). S.V.K. acknowledges the support from the Ministry of Education and Science of Russian Federation (Project 16.9790.2019). D.D.S. acknowledges the support of IUF (Institut Universitaire de France).

\end{acknowledgments}



\appendix
\counterwithin{figure}{section}

\section{Definition of IKE spectrum}

According to Ref. \cite{bradley2012energy}, the kinetic energy can be calculated for the wave function via the density weighted velocity field in space domain:

\begin{equation}
E^{(i,c)} = \frac{m}{2} \int d\mathbf{r} n(\mathbf{r})\left( |v_x^{(i,c)}(\mathbf{r})|^2 + |v_y^{(i,c)}(\mathbf{r})|^2\right),
\label{eq_ic_PureVelocityR}
\end{equation}
where $i$ and $c$ indexes correspond to incompressible and compressible velocity parts and $n=|\psi|^2$ is the density. This requires obtaining an instantaneous information on both density and phase of the condensate, which can be obtained using interferometry \cite{Sala2015}.
The equation above can be rewritten as
\begin{equation}
E^{(i,c)} = \frac{m}{2} \int d\mathbf{r} \left(|u_x^{(i,c)}(\mathbf{r})|^2 + |u_y^{(i,c)}(\mathbf{r})|^2\right),
\label{eq_ic_DWVelocityR}
\end{equation}
where the density-weighted velocity $\mathbf{u}^{(i,c)}$ is defined as follows: $\mathbf{u}^{(i,c)}=\sqrt[]{n} \mathbf{v}^{(i,c)}$. The incompressible and compressible density-weighted velocity parts should obey the following relations:

\begin{eqnarray*}
\nabla \cdot \mathbf{u}^{(i)} = 0\\
\nabla \times \mathbf{u}^{(c)} = 0
\end{eqnarray*}

These definitions straightly match with the Helmholtz decomposition of a vector field to the incompressible and compressible parts. 

Eq. \eqref{eq_ic_DWVelocityR} can be rewritten in the momentum domain 
\begin{equation}
E^{(i,c)} = \frac{m}{2} \int d\mathbf{k} \left(|u_x^{(i,c)}(\mathbf{k})|^2 + |u_y^{(i,c)}(\mathbf{k})|^2\right),
\label{eq_ic_DWVelocityK}
\end{equation}
where $\mathbf{u}^{(i,c)}(\mathbf{k})$ are the Fourier images of $\mathbf{u}^{(i,c)}(\mathbf{r})$. The Fourier components $\mathbf{u}^{(i,c)}(\mathbf{k})$ of the incompressible and compressible density-weighted velocity parts obey the following relation in the momentum domain:

\begin{eqnarray*}
\mathbf{k} \cdot \mathbf{u}^{(i)}(\mathbf{k}) = 0,\\
\mathbf{k} \times \mathbf{u}^{(c)}(\mathbf{k}) = 0.
\end{eqnarray*}

If the angular dependence is integrated out, the spectral energy density can be written as:
\begin{equation}
\label{eq_eofk}
E^{(i,c)}(k) = \frac{mk}{2} \int d\mathbf{\theta} \left( |u_x^{(i,c)}(\mathbf{k})|^2 + |u_y^{(i,c)}(\mathbf{k})|^2\right),
\end{equation}
where the absolute value $k$ and the polar angle $\theta$ define the wave vector $\mathbf{k}$. It is this spectral energy density which is expected to scale as $E^{(i)}(k) \propto k^{-5/3}$ in both the direct and the inverse energy cascades. 

\section{Numerical decomposition of the density-weighted velocity field to compressible and incompressible parts}
\label{subs_decompose}

In order to ensure that the results we obtain are not method-dependent, we have used several techniques of numerical decomposition of the velocity field into compressible and incompressible parts. We present them in this section.

(i) \textbf{The composition in the Fourier (momentum) domain}

Incompressible components can be obtained from a given density-weighted velocity field in the momentum domain $\mathbf{u}(\mathbf{k})$ as follows:

\begin{eqnarray}
u^{(i)}_{_\alpha}(\mathbf{k})=\sum_{\beta=x,y} \left( \delta_{\alpha,\beta} - \frac{k_{\alpha} k_{\beta} }{k^2} \right)u_{\beta}(\mathbf{k}), \\
u^{(c)}_{_\alpha}(\mathbf{k})=\sum_{\beta=1,2}  \frac{k_{\alpha} k_{\beta} }{k^2} u_{\beta}(\mathbf{k}),
\end{eqnarray}
where $\alpha$ and $\beta$ indices are the Cartesian coordinate directions.

(ii) \textbf{Decomposition in the spatial domain}

Here one operates fully in the spatial domain and the incompressible and compressible velocity parts are defined via the vector potential $\Phi$ and scalar potential $\phi$ as follows\cite{kowal2010velocity}:

\begin{eqnarray*}
\mathbf{u}^{(i)}(\mathbf{r}) = \nabla \times \Phi(\mathbf{r}),\\
\mathbf{u}^{(c)}(\mathbf{r}) = \nabla \cdot \phi(\mathbf{r}).\\
\end{eqnarray*}

The vector potential can be derived via

\begin{multline}
\Phi(\mathbf{r}') = \int d\mathbf{r} (\nabla \times \mathbf{u})(\mathbf{r})\cdot G(\mathbf{r}-\mathbf{r}')\\
-\oint d\mathbf{s} (\mathbf{n} \times \mathbf{u})(\mathbf{r})\cdot G(\mathbf{r}-\mathbf{r}'),
\label{eq_IntVectPot}
\end{multline}
where $G$ is the Green's function of Poisson equation in the space of a given dimensions. For the considered here 2D problem it is $G(\mathbf{r}-\mathbf{r}')=\frac{1}{2\pi} \ln (|\mathbf{r}-\mathbf{r}'|)$. In 3D one should use $G(\mathbf{r}-\mathbf{r}')=\frac{1}{4\pi |\mathbf{r}-\mathbf{r}'|}$.

For the scalar potential $\phi$ one can write in the same manner
\begin{multline}
\phi(\mathbf{r}') = \int d\mathbf{r} (\nabla \cdot \mathbf{u})(\mathbf{r})\cdot G(\mathbf{r}-\mathbf{r}')\\
-\oint d\mathbf{s} (\mathbf{n} \cdot \mathbf{u})(\mathbf{r})\cdot G(\mathbf{r}-\mathbf{r}').
\label{eq_IntScaPot}
\end{multline}
In the numerical implementation, the integrals \eqref{eq_IntVectPot} and \eqref{eq_IntScaPot} can be taken as a convolution of a matrix representing the curl of the velocity field $\nabla \times \mathbf{u}$ and a matrix for a Green's function $G(\mathbf{r})$. The latter has a size of $2N \times 2N$ with the $\mathbf{r} = 0$ corresponding to the center of the matrix: $(N,N)$ cell (here $N$ is the mesh size). 

After the incompressible $\mathbf{u}^{(i)}(\mathbf{r})$ and compressible $\mathbf{u}^{(c)}(\mathbf{r})$ parts of the density-weighted velocity are derived, one makes the Fourier transform and uses the formula \eqref{eq_eofk}. The second terms in Eqs. \eqref{eq_IntVectPot} and \eqref{eq_IntScaPot} can be omitted due to periodic boundary conditions. The same is for widely used "cup" simulations due to zero density $n$ at the boundaries.

(iii) \textbf{Mixed decomposition}

Eqs. \eqref{eq_IntVectPot} and \eqref{eq_IntScaPot} are \textit{de facto} the solution of the following spatial domain Poisson equations for the vector and scalar potentials:

\begin{eqnarray*}
\Delta \Phi = \nabla \times \mathbf{u}, \\
\Delta \phi = \nabla \cdot \mathbf{u},
\end{eqnarray*}
with the source terms being the curl and the divergence of the given velocity field $\mathbf{u}$. 
These Poisson equations can be solved using the Fourier transform scheme:

\begin{eqnarray*}
k^2 \Phi = (\nabla \times \mathbf{u})(\mathbf{k}), \\
k^2 \phi = (\nabla \cdot \mathbf{u})(\mathbf{k}),
\end{eqnarray*}
In 3D space the written above systems contains 4 equations. For the actual 2D case the curl of the field $\mathbf{u}$ aligned in ($x,y$) plane has only the $z$ component and thus one has only two equations.

\section{Analytical derivation of IKE spectra schemes}

\begin{figure}[tbp]
  \centering
  \includegraphics[width=0.49\textwidth]{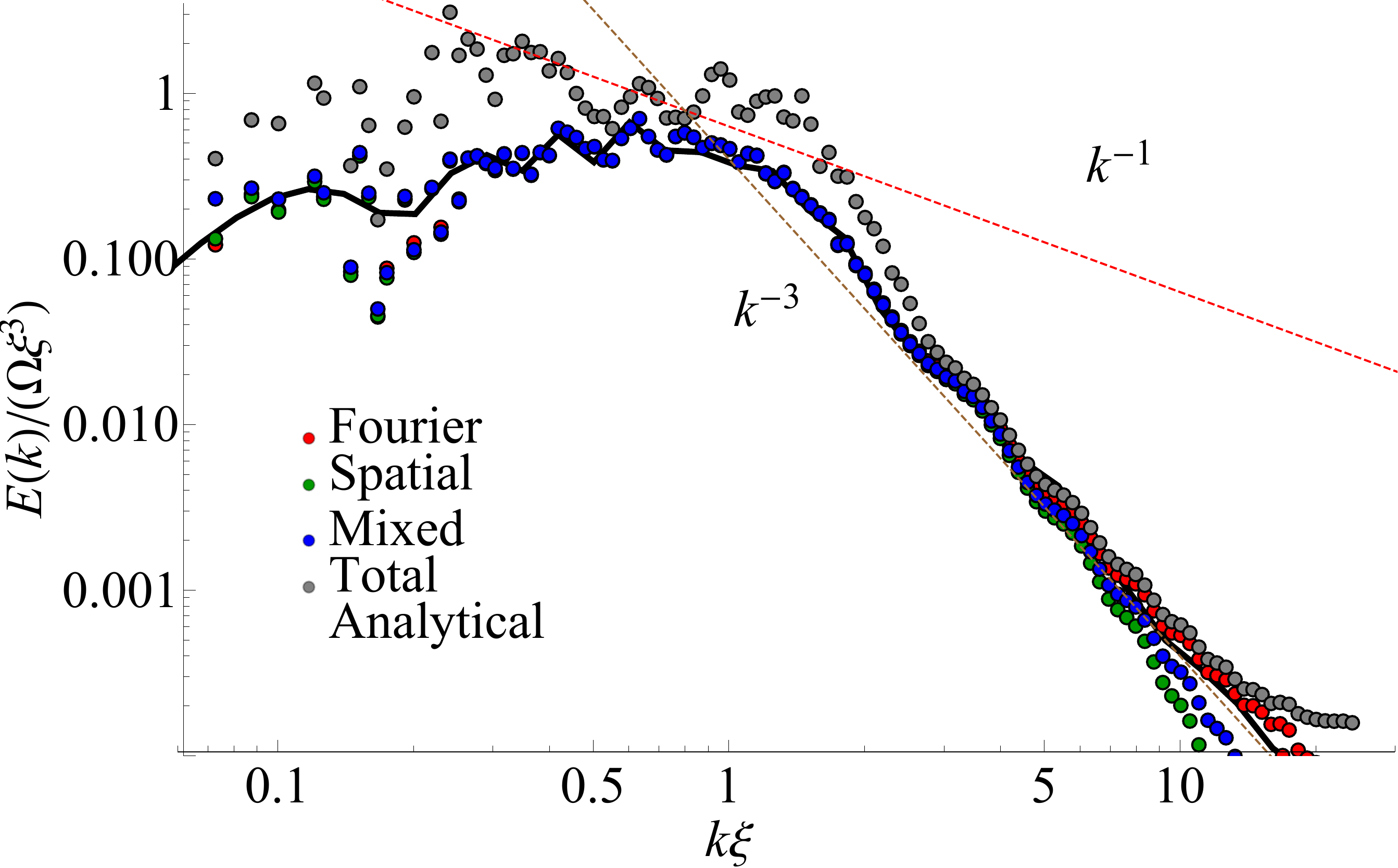}\\
  \caption{Verification of IKE spectra calculation. 3 numerical schemes: Fourier, Spatial, Mixed (points) and analytical (curve). Gray points show the total kinetic energy (mostly rotational and affected by bogolons only at low wave vectors). The mesh and the region size parameters were the same as in other simulations, but the condensate density was approx. 9 $\mu$m$^{-2}$ (healing length $\xi\approx$5.5$\mu$m, corresponding wave vector denoted with black arrow). Stirring was performed using four spoons. 8 vortices were generated.}\label{figA0}
\end{figure}

The result of the comparison of 3 numerical schemes of IKE derivation and analytical approach is given in Fig. \ref{figA0}. Unlike other figures the lower condensate density (9 $\mu$m$^{-1}$) was used to give the possibility to look closer into the vortex core.

\section{Comparison of the IKE spectra}

\begin{figure}[tbp]
  \centering
  \includegraphics[width=0.49\textwidth]{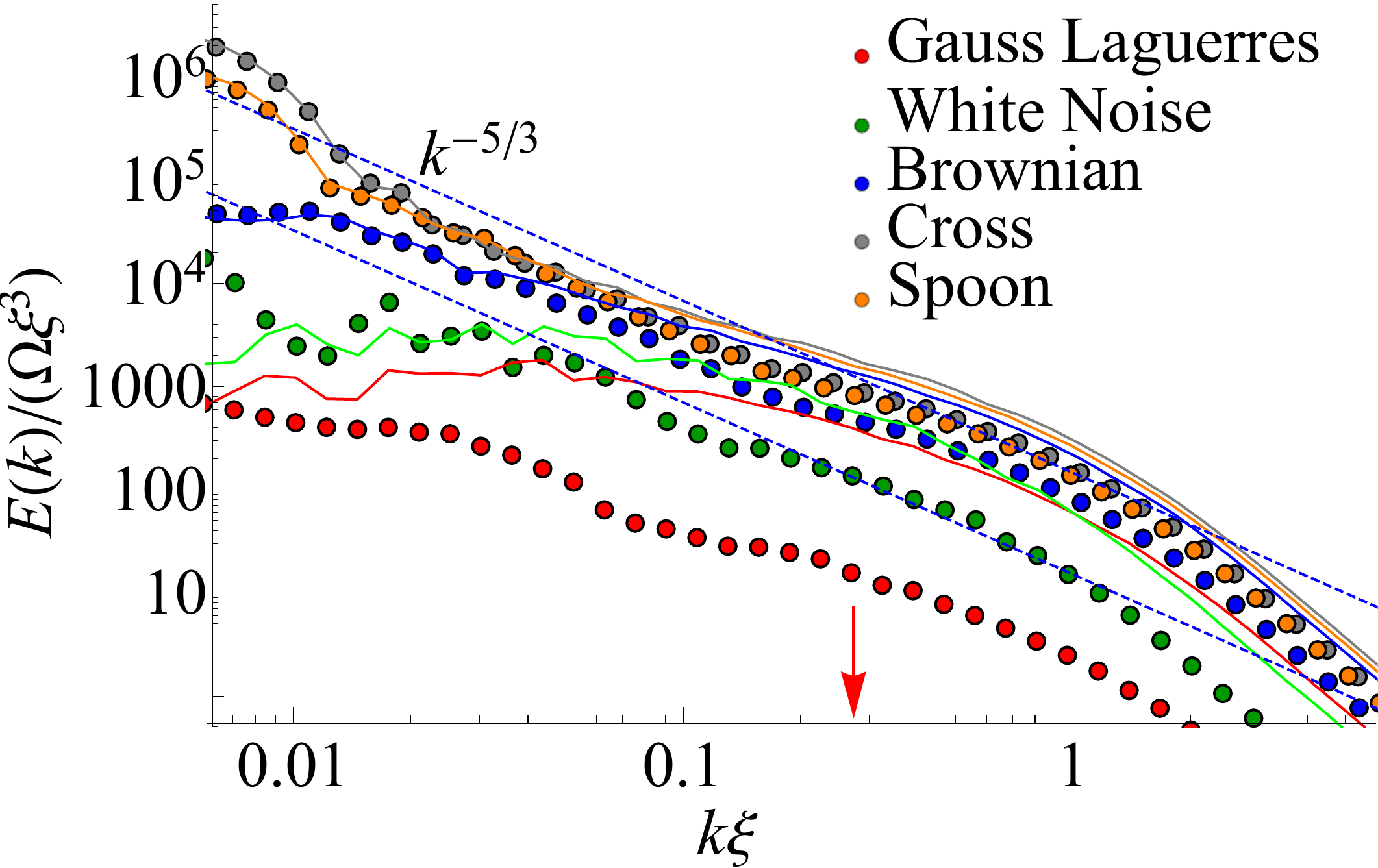}\\
  \caption{IKE spectra obtained analytically for clustered vortices only (points) and for all vortices (curves). One sees the high similarity at low wave vectors while for high wave vectors energy for clusters goes down due to "truncating" single vortices and dipoles. Red arrow denotes the inverse characteristic intervortex distance.}\label{figA2}
\end{figure}

Fig. \ref{figA2} compares the IKE spectra analytically for all vortices and for clustered vortices only. One sees that at low wave vectors the curves coincide with high accuracy. It means that the macroscopic motion of the condensate is defined by the vortex clusters only. On the contrary, at the wave vectors larger than wave vector $k_l$ the macroscopic motion can not be seen and IKE spectrum magnitude is proportional to the number of vortices only. For such stirring schemes like spoon or large cross number of clustered vortices is very large and thus IKE spectra for all vortices and for clustered vortices nearly coincide. On the contrary for GL, white noise, and Brownian spots schemes the difference is significant: the number of clustered vortices is lower.

\section{Clustering procedure}

The formation of an energy cascade is necessarily accompanied with the formation of spatial structures at different scales. For the incompressible part of the quantum fluid, it means the formation of clusters of quantum vortices of different sizes. To confirm the formation of such clusters and to separate their contribution from that of an uncorrelated vortex gas, we use the cluster detection technique described in the present section.

We follow the cluster selection procedure described in section VI of Ref. \cite{valani2018einstein}. We begin with the creating the list of vectors $\mathbf{l}_i = {(l(i,1),l(i,2), ... ,l(i,\mathrm{NOS}_i))}$, consisting of the indices of the neighbors of $i$-th vortex sorted by increasing the distance ($i=1..N_{\mathrm{vort}}$, where $N_{\mathrm{vort}}$ it the total amount of vortices). The latter (or the only) member of $\mathbf{l}_i$ is the index of the nearest vortex of the opposite sign, thus $\mathrm{NOS}_i-1$ is the number of the same-signed neighbors of $i$-th vortex lying closer than the Nearest Opposite Sign neighbor. After that, we create the $\mathbf{l}'_i$ vectors by dropping the last element and thus $\mathbf{l}'_i$ vectors list only the neighbors of the same sign of $i$-th vortex. In some cases (e.g. for vortex belonging to the dipole) $\mathbf{l}'_i$ can be empty.

At the first step, we find the vortex pairs by finding the pairs of $i$ and $j$ indices so that $\mathrm{NOS}_i = 1$ and $\mathrm{NOS}_{l(i,1)}=1$ is also equal to one. After that the pair $(i,l(i,1))$ is put to the list of connected vortices $\mathbf{L}$.

At the second step, for all $i$ and for all $j \leq \mathrm{NOS}_i-1$ we put the pair $(i,l'(i,j))$ to the list of connected vortices $\mathbf{L}$ if $\mathbf{l}'_{l(i,j)}$ contains index $i$. This procedure is in fact the finding of mutual vortex pairings.

Then we consider the list of connected vortices $\mathbf{L}$ (containing both vortex pairs and the clusters of the same-sign vortices) as a graph and separate its connectivity components using the \texttt{ConnectedGraphComponents} routine of the \textsc{Mahtematica} package. Each connectivity component is thus a cluster. If the connectivity component consist of exactly two vortices with opposite circulation, it is marked as a pair, otherwise it is counted as a cluster.

\section{Box-counting fractal dimension}

\begin{figure}[tbp]
  \centering
  \includegraphics[width=0.23\textwidth]{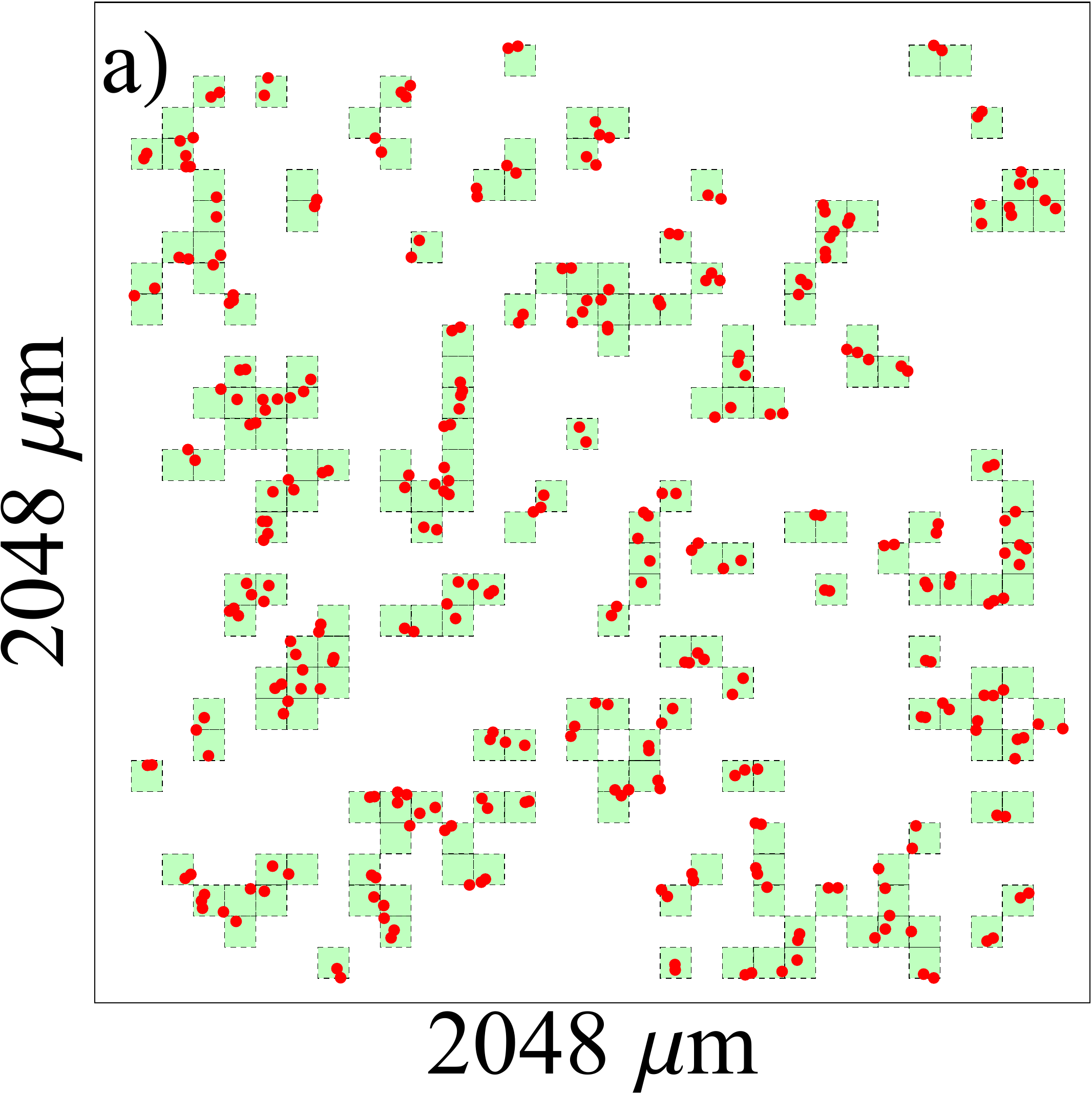}\hspace{10pt}\includegraphics[width=0.23\textwidth]{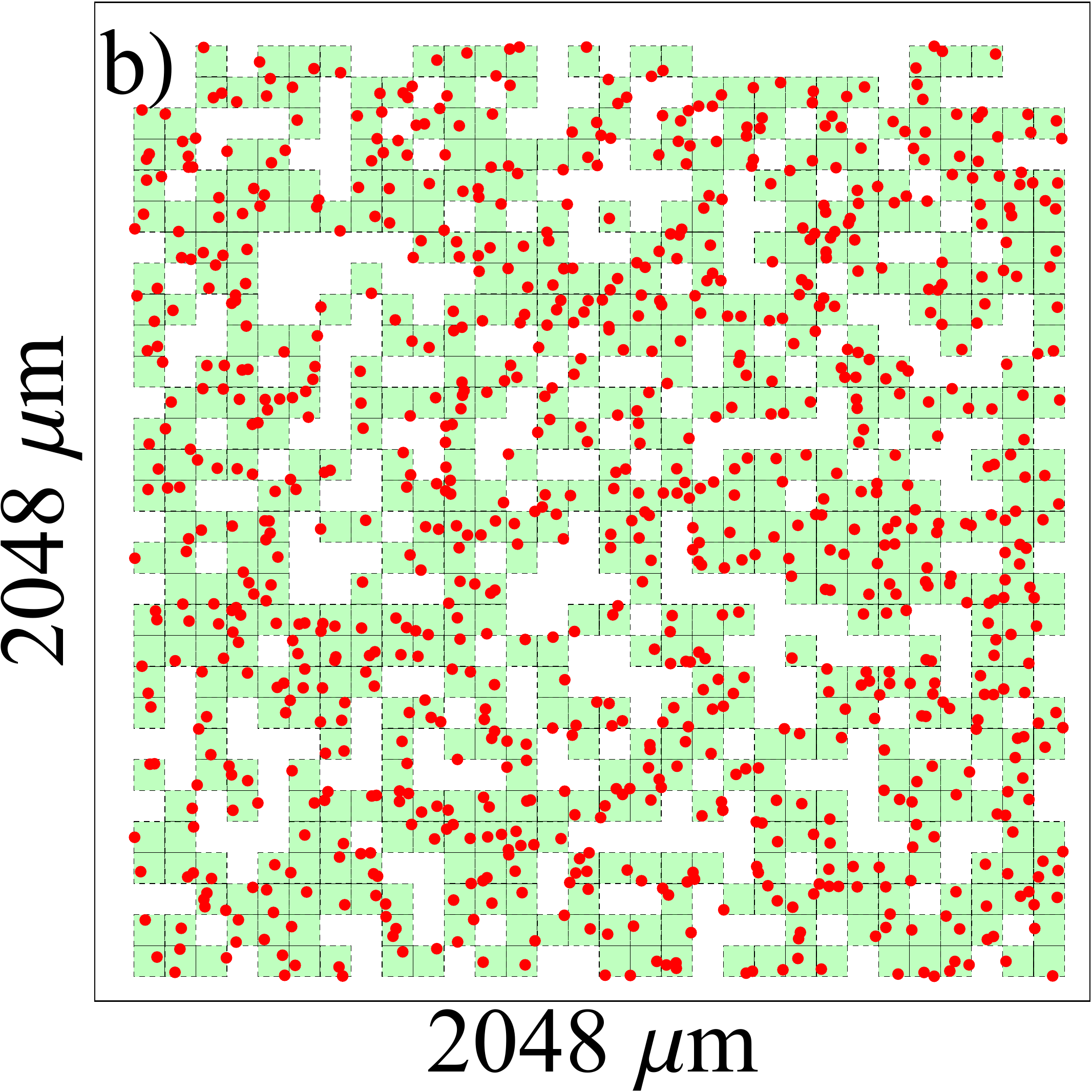}
  
  \caption{Illustration of the box counting algorithm applied to determine the fractal dimension of the patterns formed by votrtices. Box size corresponds to $k\xi=0.9$. Each vortex is denoted with a red point. Only the same-sign vortices were taken into account. Panel a) is for clustered vortices only and panel b) is for all vortices.}\label{fig_App_frac}
\end{figure}

Figure~\ref{fig_App_frac} illustrates the box-counting algorithm that we have used for the determination of the fractal dimension of the structures formed by vortices. For this, we considered the clusters of vortices of the same sign (for the opposite sign the picture is the same). The box size shown in the figure is chosen as an example of the transitional regime $2\pi \xi/\varepsilon=0.9$. For smaller box sizes, each vortex is covered with a single box and the system is effectively 0D. On the contrary, for larger $\varepsilon$, all system is covered by boxes and thus it is 2D. The spatial distribution of all vortices without cluster selection exhibits the 2D nature already at this size scale and does not exhibit an evident transitional regime.

To prove that the transitional region observed for clustered vortices indeed corresponds to what one would expect for a fractal structure, we have compared it with a set of randomly-distributed points and with a well-known fractal structure (Sierpinski triangle). The latter was generated by the so called chaos game method. Starting from a randomly chosen point $\mathbf{v}_1$ in the triangle with the vertices $\mathbf{p}_1$, $\mathbf{p}_2$, and $\mathbf{p}_3$, one consequentially makes the steps directed to randomly chosen triangle vertex but passing only half of required distance. The corresponding recurrent formula reads $\mathbf{v}_{i+1} = (\mathbf{v}_i + \mathbf{p}_{r_i})/2$, where $r_i$ a random integer from 1 to 3. Such Sierpinski triangle-like patterns were generated for 320  points and then rescaled to give the same mean distance between the points as the one observed in the vortex distributions (approx. $25\xi$). Finally, all space was tiled with a 2D lattice of such patterns to obey the transition to 2D regime at large scales. 

\begin{figure}[tbp]
  \centering
  \includegraphics[width=0.5\textwidth]{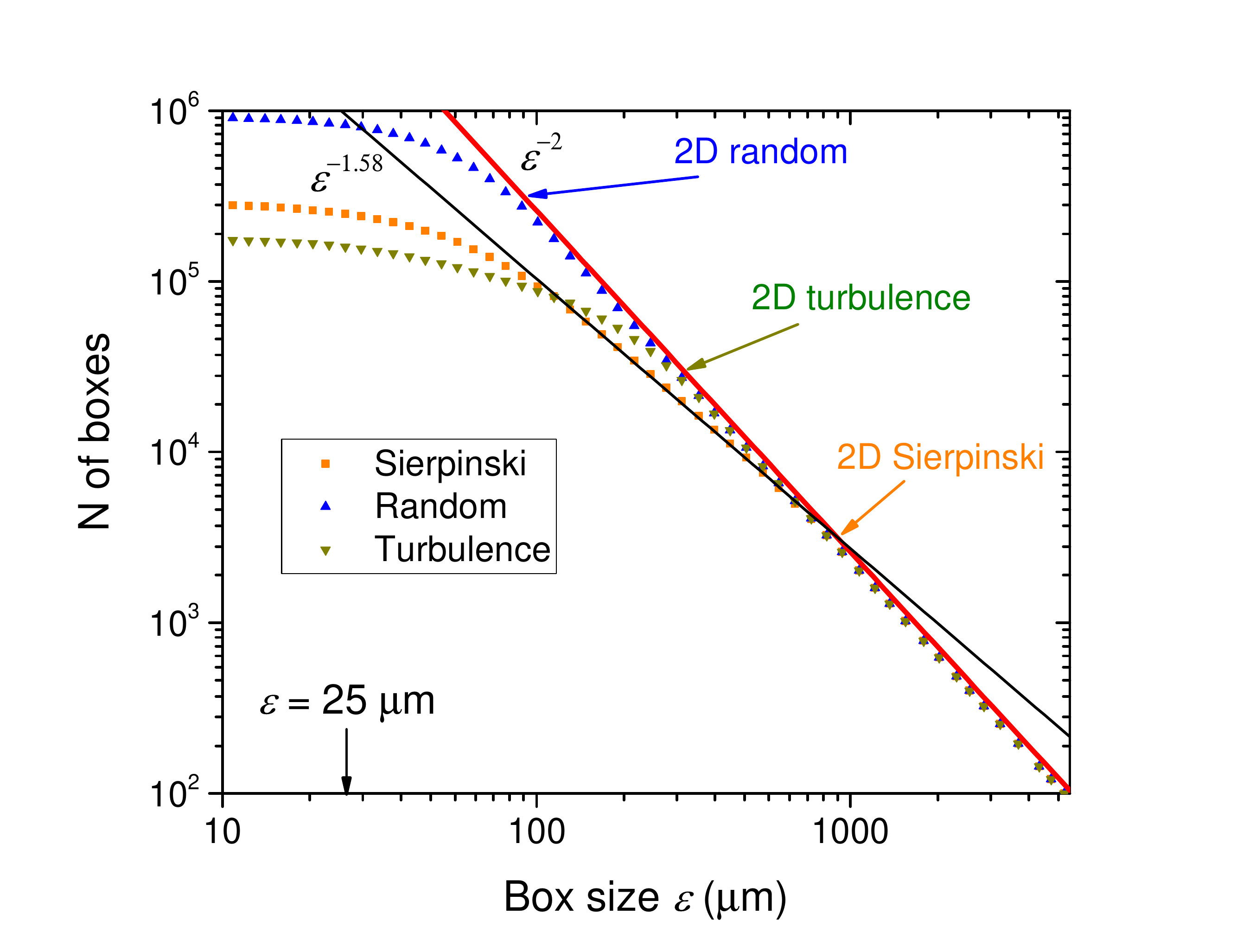}\\
  \caption{Comparison of vortex clustering algorithm for clusters of vortices and artificially generated patterns in the shape of Sierpinski triangle and random spatial distribution. The power function of $\log_23 \approx 1.58$ corresponds to the exact value of Sierpinski triangle fractal dimension.}\label{fig_App_Sierp}
\end{figure}

Figure~\ref{fig_App_Sierp} allows to compare the size of the transitional region for randomly distributed points (blue), vortex structures arising from the turbulence (green), and a perfect fractal structure of the Sierpinski triangle (orange). For random points, this region is the smallest and no fractal dimension can be determined. For the vortex clusters and the Sierpinski triangle, the transitional region is much larger, and a fractal dimension of $\log_23$ can be determined correctly for the Sierpinski triangle. We conclude therefore that vortex clusters indeed form a fractal structure. The fact that the size of the transitional region for the vortex clusters is slightly smaller than for the triangle could be partially explained by the fact that the fractal dimensions are different, and the transition to the 2D exponent is therefore smoother in the case of the Sierpinski fractal.

At the same time, Fig.~\ref{fig_App_Sierp} shows clearly that in a finite-size system the fractal dimension region does not have an infinite extension even for a perfect self-similar distribution of points. In order to robustly observe the intermediate regime with fractional Minkowski-Bouligand dimension one requires the system size to be at least two orders higher than the average distance between the points. Practically, for polaritons such system sizes of several hundreds of microns are already achievable, and increasing them to the scale of $1$~mm should allow to significantly increase the reliability of the determination of the fractal dimension.
\section{Parameters of the stirrers}

1. \textit{Large cross-like potential}\\
The length of the cross was 860 $\mu$m and the width 100 $\mu$m (with additional 64 $\mu$m Gaussian filter-based smoothing of borders). The full 360$^{\circ}$ rotation took 1280 ps. The potential depth $V=10$ meV. Duration of stirring was 1.5 ns and total simulation time was 25 ns. 

2. \textit{Gauss-Laguerres}\\

The stirring was performed by 32 randomly placed rotating potentials during first 0.5 ns (total simulation time was 25 ns). The potential depth $V=10$ meV. Profile of each stirrer was given by a superposition of the two 2$^{\mathrm{nd}}$ order Gauss-Laguerre beams. One of them was stationary and the second one was rotating. To obtain the potential profile, the electric field magnitude square was taken. The resulting profile resembled the 4 smoothed spots with the distance between the opposite ones 20 $\mu$m, see also the sketch in Fig. \ref{fig1_stirring}. The full 360$^{\circ}$ rotation of resulting potential (not the electric field) took 45 ps, which yields approximately the same linear velocity as for large cross strategy.

3. \textit{Classical rotating spoon}\\
The orbit diameter of the spoon was 632 $\mu$m. The shape of the spoon was given by $\tilde{V}(r)=(\exp((r-32\mu \mathrm{m})/(2.5 \mu\mathrm{m}))+1)^{-1}$. The full 360$^{\circ}$ rotation took 1280 ps as for the cross. The potential depth $V=10$ meV. Duration of stirring was 3.0 ns and total simulation time was 25 ns.

4. \textit{Several spots in Brownian motion}\\
The trajectory was obtained as a Beta Spline curve defined by the points obtained by random walks. Distance of each step was fixed to 50 $\mu$m and the direction was random (30 ps between two steps). Hence, the speed of the spots was approx 2 $\mu$m/ps. Number of spots was 10, total simulation time 50 ns, stirring time 1.5 ns. The spots were in the shape of Gaussian profiles with the radii of $7~\mu$m and the potential depth $V=10$ meV. The routine \textsc{BSplineFunction} in the \textsc{Mathematica} package was used to obtain the curve.

5. \textit{White noise}\\
White noise was obtained from 1024x1024 matrix of uniform random values from 0 to 1 multiplied by the amplitude 80 meV. Then for smoothing and thus bringing some finite spatial correlations the Fourier image was filtered with the Gaussian function in reciprocal space. The width was of Gaussian was $\frac{2\pi}{r_{\mathrm{correl}}}$, where the correlation length $r_{\mathrm{correl}}=75~\mu \mathrm{m}$. Instantaneous switching the potential to the new random realization was performed each 0.4 ps. Total simulation time was 50 ns and the white noise potential was applied during the first 5 ns.

\bibliography{bib}

\end{document}